\documentclass[aps,prb,showpacs,twocolumn]{revtex4-1}
\usepackage{amsmath,amssymb,bm,dcolumn,epsfig,graphicx,longtable}
\begin{document}

\title{Simple way to apply nonlocal van der Waals functionals within all-electron methods}
\author{Fabien Tran}
\author{Julia Stelzl}
\author{David Koller}
\author{Thomas Ruh}
\author{Peter Blaha}
\affiliation{Institute of Materials Chemistry, Vienna University of Technology,
Getreidemarkt 9/165-TC, A-1060 Vienna, Austria}

\begin{abstract}

The method based on fast Fourier transforms proposed by
G. Rom\'{a}n-P\'{e}rez and J. M. Soler [Phys. Rev. Lett. \textbf{103}, 096102 (2009)],
which allows for a computationally fast implementation
of the nonlocal van der Waals (vdW) functionals, has significantly contributed
to making the vdW functionals popular in solid-state physics. However, the
Rom\'{a}n-P\'{e}rez-Soler method relies on a plane-wave expansion of the
electron density; therefore it can not be applied readily to all-electron
densities for which an unaffordable number of plane waves would be required
for an accurate expansion. In this work, we present the results for the lattice
constant and binding energy of solids that were obtained by applying a
smoothing procedure to the all-electron density calculated with the
linearized augmented plane-wave method. The smoothing procedure has
the advantages of being very simple to implement, basis-set independent, and
allowing the calculation of the potential.
It is also shown that the results
agree very well with those from the literature that were obtained with
the projector augmented wave method.

\end{abstract}

\maketitle

\section{\label{introduction}Introduction}

It is well known that in density functional theory (DFT)
\cite{HohenbergPR64,KohnPR65} the exchange-correlation (xc) functionals
$E_{\text{xc}}$ of the semilocal and hybrid approximations do not account
properly for the van der Waals (vdW) interactions since the attractive part,
the London dispersion forces, is missing and, as a consequence,
lead to erratic results when applied to systems where the vdW interactions
play an important role.\cite{GrimmeCR16} In this respect, two extreme and
opposite functionals are the local density
approximation (LDA)\cite{KohnPR65} and the generalized gradient approximation
(GGA) of Becke, Lee, Yang, and Parr \cite{BeckePRA88,LeePRB88} that lead to severe
overbinding and underbinding (or even no binding at all),
respectively.\cite{KristyanCPL94,PerezJordaCPL95} Therefore, dispersion
correction terms to be added to a semilocal (SL) or hybrid xc functional,
\begin{equation}
E_{\text{xc}} = E_{\text{xc}}^{\text{SL/hybrid}} + E_{\text{c,disp}},
\label{Exc}
\end{equation}
have been proposed, such that much more reliable results can be obtained for
vdW systems with DFT (see Refs.~\onlinecite{KlimesJCP12,DobsonIJQC14,BerlandRPP15,GrimmeCR16} for
reviews).

The most simple and popular type of correction for
$E_{\text{c,disp}}$ consists of an
atom-pairwise term\cite{WuJCP01,WuJCP02,HasegawaPRB04,GrimmeJCC04}
for each atom pair $A$-$B$ (separated by $R_{AB}$):
$-f^{\text{damp}}\left(R_{AB}\right)C_{6}^{AB}/R_{AB}^{6}$,
where $C_{6}^{AB}$ are the dispersion coefficients and $f^{\text{damp}}$ is
a damping function. Such corrections lead to little increase in the
computational time and have proven to be very useful for improving the
reliability of DFT calculations.\cite{GrimmeCR16} Depending on the particular scheme,
the coefficients $C_{6}^{AB}$ are precomputed
\cite{WuJCP01,WuJCP02,HasegawaPRB04,GrimmeJCC04} or evaluated on the fly by
using properties of the system like the atomic positions or the electron
density.\cite{BeckeJCP05,TkatchenkoPRL09,GrimmeJCP10}
Nowadays, the most used atom-pairwise methods are DFT-D3 from Grimme and
co-workers \cite{GrimmeJCP10,GrimmeJCC11} and PBE+TS from Tkatchenko
and Scheffler.\cite{TkatchenkoPRL09,TkatchenkoPRL12} The atom-pairwise methods
can be applied to molecules and extended systems.

Also popular, are the so-called nonlocal (NL) vdW dispersion terms of the
form\cite{DionPRL04,LangrethJPCM09}
\begin{equation}
E_{\text{c,disp}}^{\text{NL}} = \frac{1}{2}\int\int\rho(\bm{r}_{1})
\Phi\left(\bm{r}_{1},\bm{r}_{2}\right)\rho(\bm{r}_{2})d^{3}r_{1}d^{3}r_{2},
\label{EcdispNL}
\end{equation}
where the kernel $\Phi$ depends on the electron density $\rho$ and its derivative
$\nabla\rho$ as well as on $\left\vert\bm{r}_{1}-\bm{r}_{2}\right\vert$.
The first functional of the form of Eq.~(\ref{EcdispNL}) that could be applied
to all kinds of system was proposed by
Dion \textit{et al}.\cite{DionPRL04} (DRSLL) and was derived as a simplification of
the adiabatic connection formula.\cite{GunnarssonPRB76,LangrethPRB77}
The DRSLL term was originally added to the semilocal functional
consisting of the GGA revPBE\cite{ZhangPRL98} for exchange and LDA
\cite{VoskoCJP80,PerdewPRB92a} for correlation.
Since then, other kernels $\Phi$ in Eq.~(\ref{EcdispNL}) or associated
semilocal functionals have been proposed by various authors.
\cite{LeePRB10,KlimesJPCM10,CooperPRB10,VydrovPRL09,VydrovJCP10,WellendorffTC10,SabatiniPRB13,HamadaPRB14,BerlandPRB14,PengPRX16}
Overall, the most recent versions of nonlocal vdW functionals and pair-wise methods
are rather similar in terms of accuracy (see, e.g.,
Refs. \onlinecite{GrimmeWCMS11,GoerigkJCT10,RegoJPCM15,TranJCP16,LozanoPCCP17}),
nevertheless, the double integration in Eq.~(\ref{EcdispNL}) makes the calculations with
the nonlocal vdW functionals more expensive.

Brute-force methods to carry out the double integration in
Eq.~(\ref{EcdispNL}) have been proposed,\cite{LazicCPC10,NabokCPC11} however
an important step in the development of nonlocal vdW functionals for periodic
systems was made by Rom\'{a}n-P\'{e}rez and Soler\cite{RomanPerezPRL09} (RPS) who
proposed a very efficient method for evaluating Eq.~(\ref{EcdispNL}).
Their method, which is based on fast Fourier transforms (FFT)
and the convolution theorem, is now the standard method
and is implemented in various solid-state codes.
\cite{SabatiniJPCM12,WellendorffTC10,KlimesPRB11,TranJCP13,LarsenMSMSE17}
Furthermore, Rom\'{a}n-P\'{e}rez and Soler also showed that their method
allows for a straightforward calculation of the functional
derivative of Eq.~(\ref{EcdispNL}) that is much simpler
than the complicated ways from Refs.~\onlinecite{ThonhauserPRB07,GulansPRB09}.
Also, in Ref.~\onlinecite{SabatiniJPCM12},
the formula for the contribution to the stress tensor was derived.
For these reasons, the RPS method is the preferred one compared
to the others that have been proposed.\cite{GulansPRB09,LazicCPC10,NabokCPC11}

However, since the RPS method relies on a plane-wave expansion of $\rho$,
it is not obvious how to apply it to all-electron densities, since their large
variations close to the nuclei would require an unrealistically large plane-wave
expansion. This is the reason why, to our knowledge,
the RPS method has been implemented only
into codes relying exclusively on a plane-wave
expansion of $\rho$, i.e., pseudopotential codes.
In order to use the RPS method within an all-electron code,
a simple solution would be to smooth the electron density close to the
nuclei, such that a pure plane-wave expansion of $\rho$ is possible.

The goal of this work is to present a smoothing procedure to
all-electron densities and to study in detail which degree of smoothing should be
applied in order to make the calculations affordable without sacrificing too
much accuracy in the calculation of Eq.~(\ref{EcdispNL}).
This is done in the framework of the linearized-augmented
plane-wave\cite{AndersenPRB75,Singh} (LAPW) method, which provides very
accurate all-electron densities. However, our smoothing method is
basis-set independent and, therefore, can be implemented in any
all-electron code. Thus, we expect our work to contribute to a much more
widespread use of the nonlocal vdW functionals in the all-electron solid-state
community.

The properties that will be considered for our tests are the lattice constant and binding
energy of various types of solids, and the results will be compared to results
that were obtained with the projector augmented-wave (PAW) method.\cite{BlochlPRB94b}

The paper is organized as follows. Section~\ref{methodology} gives details
about the used methods. Then, the results are presented and discussed in
Sec.~\ref{results}, while Sec.~\ref{summary} summarizes the main conclusions
of this work.

\section{\label{methodology}Methodology}

We start with a brief introduction to the LAPW method that is used in the
present work to solve the Kohn-Sham\cite{KohnPR65} (KS) equations of DFT.
Adopting a notation that is common to all flavors of the LAPW method,
\cite{Singh,SjostedtSSC00,MichalicekCPC13} the Bloch orbitals ($n$ is the band index
and $\bm{k}$ is a vector in the first Brillouin zone) are expanded as
\begin{equation}
\psi_{n\bm{k}}(\bm{r}) =
\sum\limits_{\bm{K}}
c_{n\bm{k}\bm{K}}
\phi_{\bm{k}\bm{K}}(\bm{r}) +
\sum\limits_{i}
c_{n\bm{k}i}^{\text{LO}}
\phi_{i}^{\text{LO}}(\bm{r}),
\label{psink}
\end{equation}
where
\begin{equation}
\phi_{\bm{k}\bm{K}}(\bm{r}) =
\left\{
\begin{array}{l@{\quad}l}
\sum\limits_{\ell,m}\sum\limits_{f}
d_{f\bm{k}\bm{K}}^{\alpha\ell m}
D_{f}^{\alpha\ell}(r_{\alpha})
Y_{\ell m}(\hat{\bm{r}}_{\alpha}), &
\bm{r}\in\text{S}_{\alpha} \\
\frac{1}{\sqrt{\Omega}}
e^{i\left(\bm{k}+\bm{K}\right)\cdot\bm{r}}, & \bm{r}\in\text{I}
\end{array}
\right.
\label{phikk}
\end{equation}
are basis functions (indexed with the reciprocal lattice vector $\bm{K}$)
that consist of a linear combination of products
between a radial function $D_{f}^{\alpha\ell}$ and a spherical harmonics
$Y_{\ell m}$ inside the atomic sphere $\text{S}_{\alpha}$ centered
at nucleus $\alpha$,
and of a plane wave in the interstitial region I. The coefficients
$d_{f\bm{k}\bm{K}}^{\alpha\ell m}$ are determined from the requirement
of continuity of $\phi_{\bm{k}\bm{K}}$ [and its derivative(s) depending on the
LAPW flavor] at the sphere boundary. The basis functions
\begin{equation}
\phi_{i}^{\text{LO}}(\bm{r}) =
\left\{
\begin{array}{l@{\quad}l}
\sum\limits_{f}
d_{\text{LO},if}^{\alpha_{i}\ell_{i}m_{i}}
D_{if}^{\alpha_{i}\ell_{i}}(r_{\alpha_{i}})
Y_{\ell_{i}m_{i}}(\hat{\bm{r}}_{\alpha_{i}}), &
\bm{r}\in\text{S}_{\alpha_{i}} \\
0, & \bm{r}\in\text{I}
\end{array}
\right.,
\label{phiLO}
\end{equation}
are the so-called local orbitals (LO) that are defined only inside the spheres
and set to zero in the interstitial region.\cite{Singh}
The number of basis functions in Eq.~(\ref{psink}) is determined by
the cutoff $K_{\text{max}}$ such that
$\left\vert\bm{k}+\bm{K}\right\vert\leqslant K_{\text{max}}$
and the number of LO in the second term.
Note that the basis-set expansion
Eq.~(\ref{psink}) is used for the valence and unoccupied orbitals, but not for the
deep-lying core orbitals which are calculated by solving the radial
KS equations inside the MT spheres numerically
without basis-set expansion.\cite{Singh}

Once the orbitals are obtained, the electron density $\rho$ is calculated
and expanded in (real) spherical harmonics and plane waves inside the atomic
spheres and interstitial regions, respectively. The great advantage of the
LAPW method is to provide a virtually exact all-electron solution of the KS
equations for a given xc functional.

On the other hand, however, the LAPW method does not allow for a straightforward use of
the RPS method to calculate Eq.~(\ref{EcdispNL}) since a pure plane-wave
expansion of the all-electron density $\rho$ is out of reach. In order to use the
RPS method with the LAPW method, the most obvious choice is to smooth $\rho$ in
the core region such that a plane-wave expansion is possible.
A simple procedure to construct a smooth density $\rho_{\text{s}}$ is given by
\begin{equation}
\rho_{\text{s}}(\bm{r}) =
\left\{
\begin{array}{l@{\quad}l}
\rho(\bm{r}), &
\rho(\bm{r}) \leqslant \rho_{\text{c}} \\
\frac{\rho(\bm{r})+A\rho_{\text{c}}\left(\rho(\bm{r})-\rho_{\text{c}}\right)^{n}}
{1 + A\left(\rho(\bm{r})-\rho_{\text{c}}\right)^{n}}, & \rho(\bm{r}) > \rho_{\text{c}}
\end{array},
\right.
\label{rhos}
\end{equation}
where $\rho_{\text{c}}$ and $n$ are parameters, and $A$ is set to 1~Bohr$^{3n}$
and is introduced only for the sake of consistency of the units [e.g., $A$
would be $(0.529177)^{3n}$~\AA$^{3n}$ if $\rho$ was expressed in~\AA$^{-3}$].
$\rho_{\text{c}}$ is the density cutoff that has to be chosen carefully such
that the plane-wave expansion of $\rho_{\text{s}}$ everywhere in the unit
cell is small enough to make the calculation with the RPS method affordable,
but without loosing too much accuracy with respect to the reference calculation with the
original density. $n$ determines until which order the derivatives
of $\rho_{\text{s}}$ are continuous, and in Appendix
\ref{appendixA} it is shown explicitly that $\nabla\rho_{\text{s}}$ and
$\nabla^{2}\rho_{\text{s}}$ are continuous if $n\geqslant2$.

\begin{figure}
\includegraphics[width=\columnwidth]{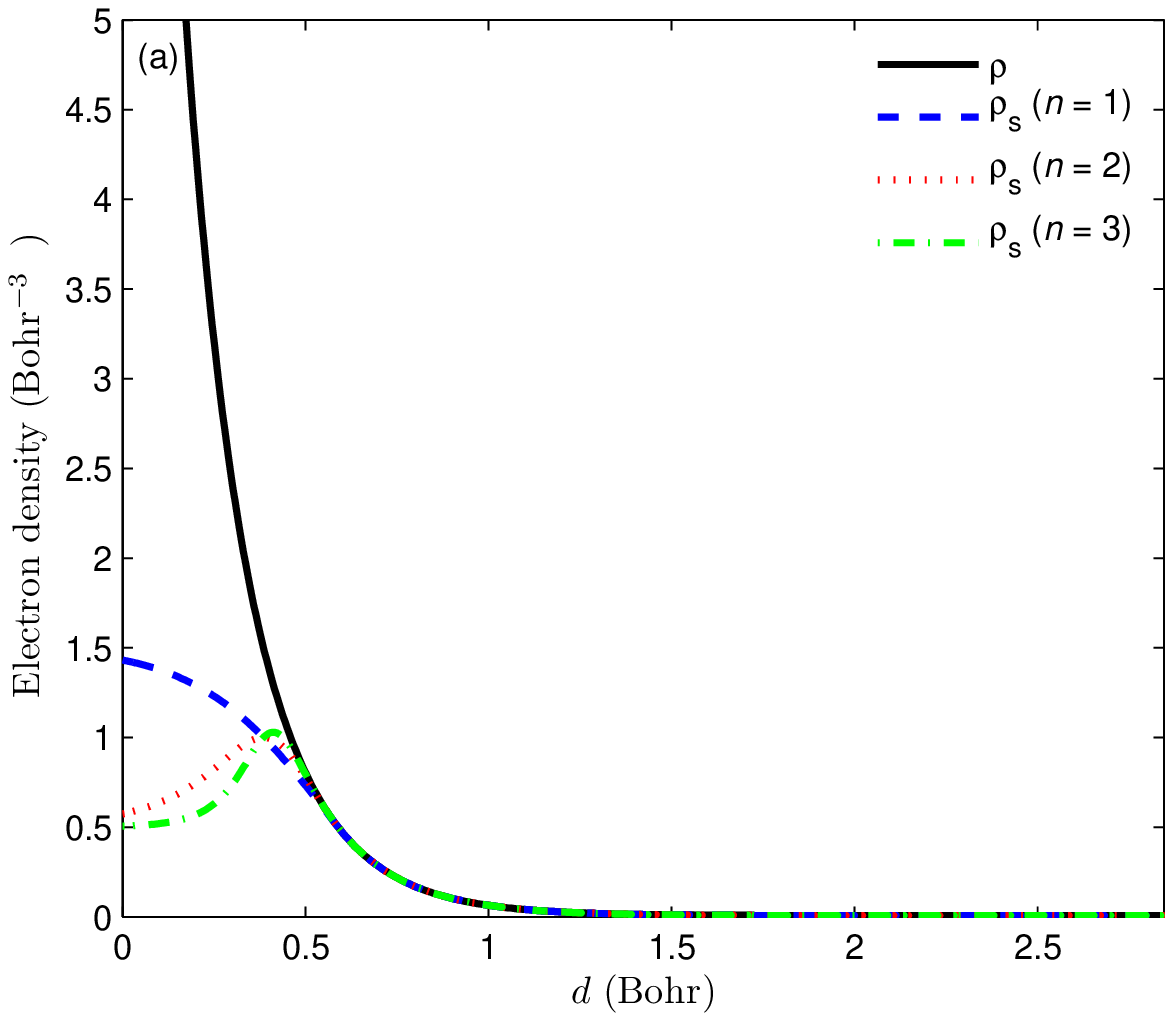}
\includegraphics[width=\columnwidth]{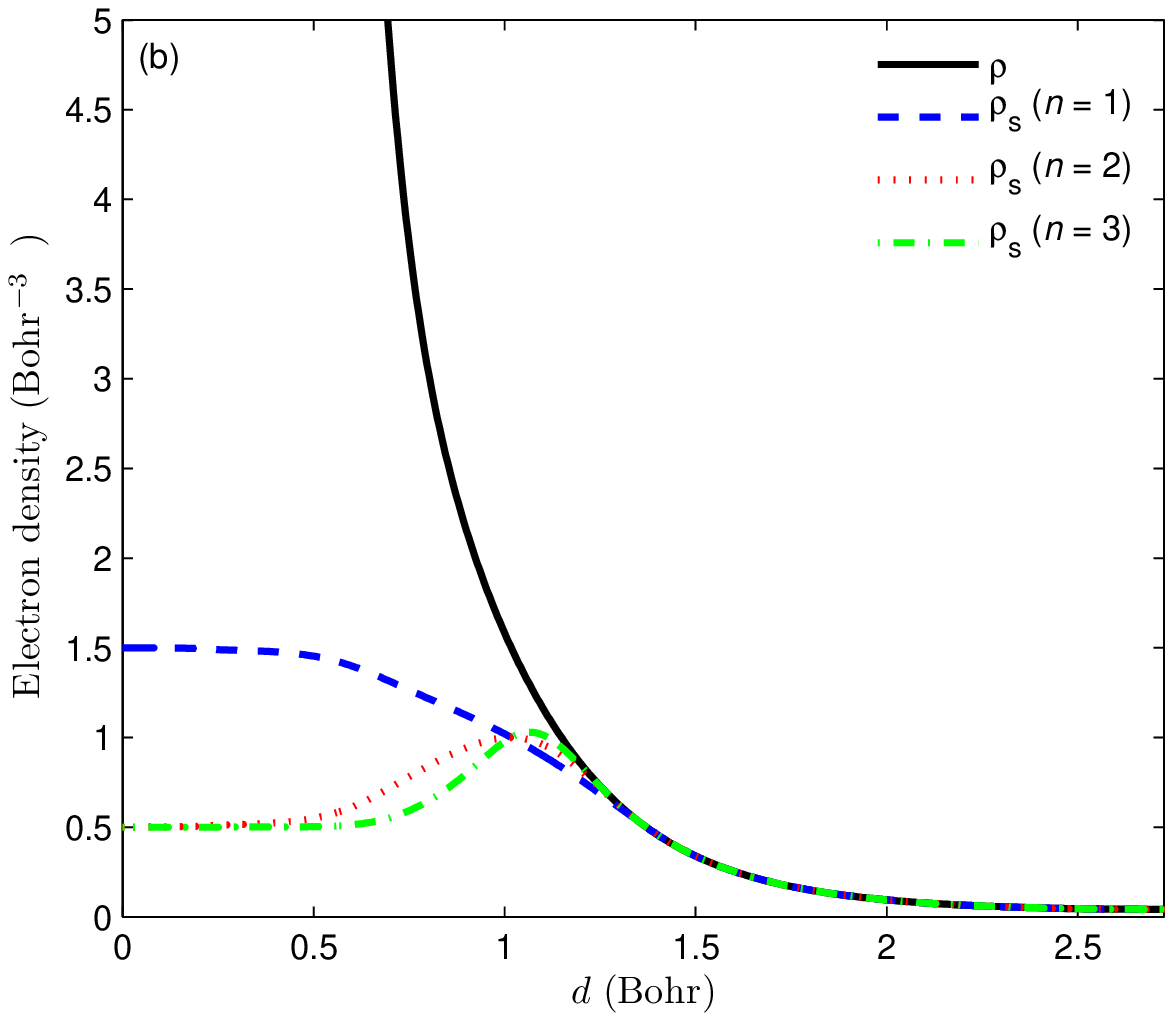}
\caption{\label{fig_rhos_1}Electron density in (a) Li and (b) Au plotted from
an atom at $d=0$ to the mid-distance to the nearest neighbor atom.
$\rho$ is the original all-electron density and $\rho_{\text{s}}$ are smooth
densities calculated from Eq.~(\ref{rhos}) for different values of $n$
with $\rho_{\text{c}}=0.5$ Bohr$^{-3}$.}
\end{figure}
\begin{figure}
\includegraphics[width=\columnwidth]{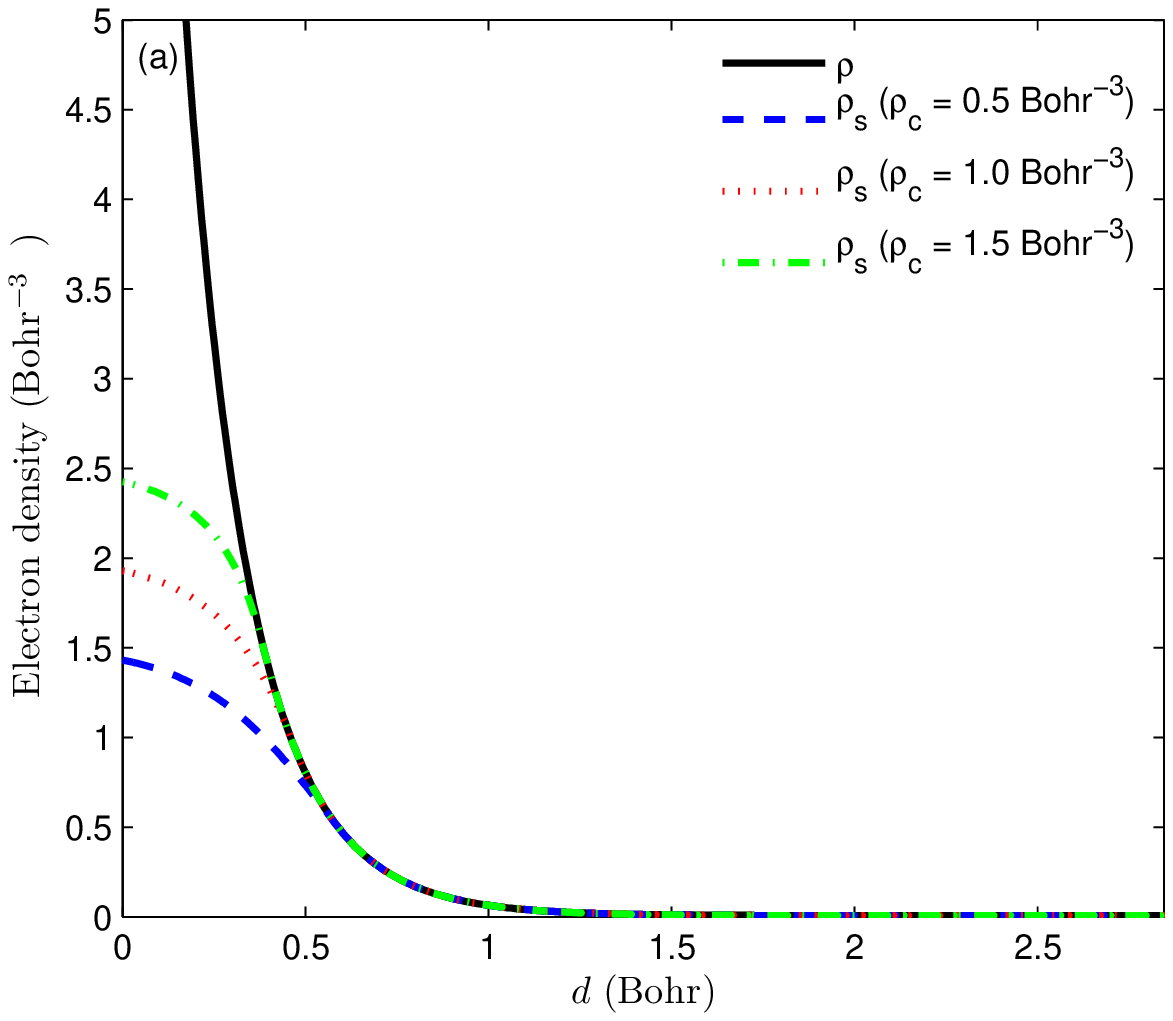}
\includegraphics[width=\columnwidth]{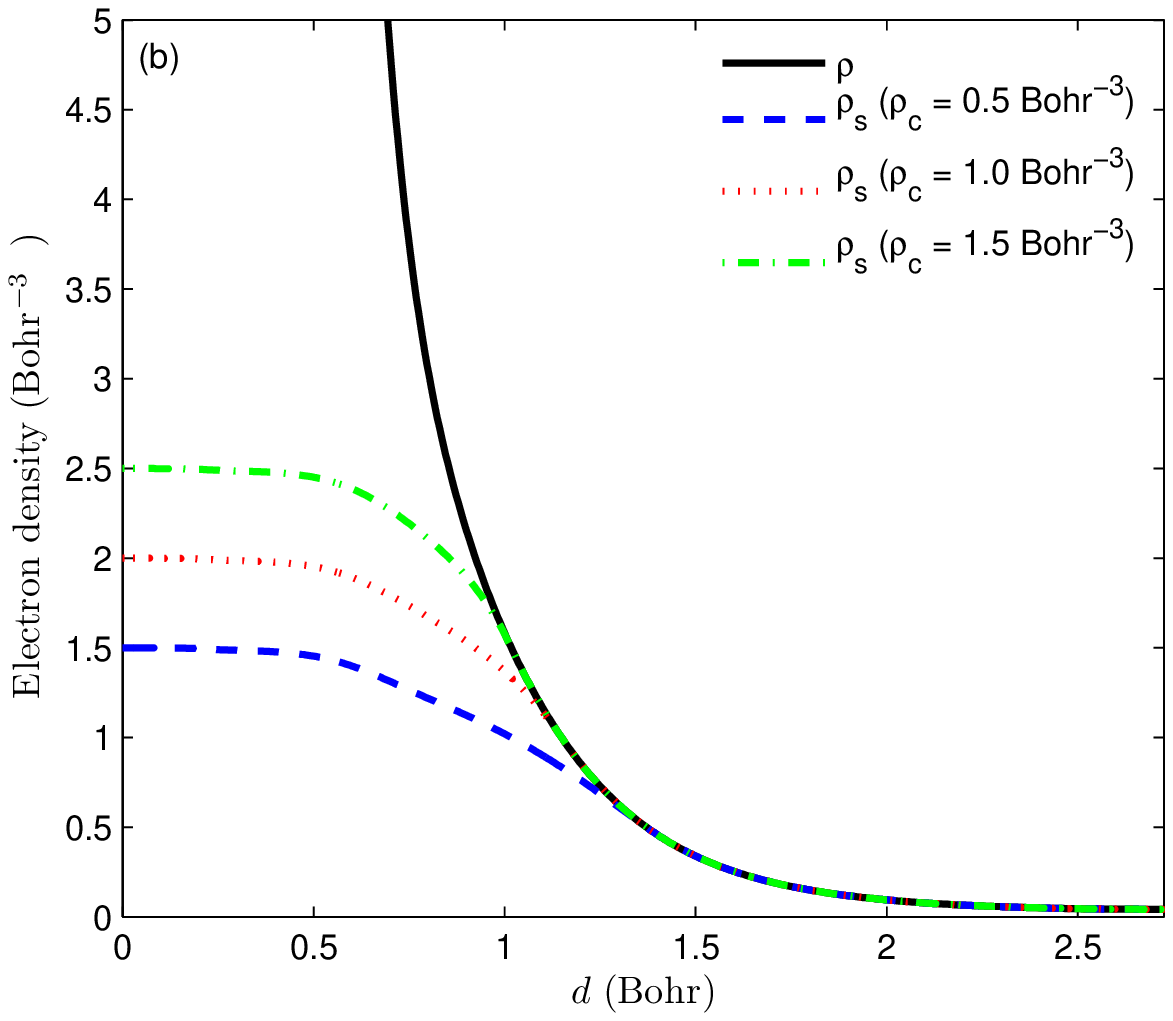}
\caption{\label{fig_rhos_2}Electron density in (a) Li and (b) Au plotted from
an atom at $d=0$ to the mid-distance to the nearest neighbor atom.
$\rho$ is the original all-electron density and $\rho_{\text{s}}$ are smooth
densities calculated from Eq.~(\ref{rhos}) for different values of $\rho_{\text{c}}$
with $n=1$.}
\end{figure}
Figures \ref{fig_rhos_1} and \ref{fig_rhos_2} show the
effect of the smoothing procedure for different values of $n$ and
$\rho_{\text{c}}$ in solid Li and Au, which represent cases of
a very light and a very heavy atom, respectively.
It is clear that a pure plane-wave expansion of such densities
$\rho_{\text{s}}$ without the peaks close to the nuclei should converge much
faster than the original density $\rho$. From Fig.~\ref{fig_rhos_1},
we can see that the smoothest curve is obtained with $n=1$ (continuity of
$\nabla\rho_{\text{s}}$), while requiring higher continuous derivatives of
$\rho_{\text{s}}$ leads to a maximum and overall to a
less smooth density $\rho_{\text{s}}$. Convergence tests of Eq.~(\ref{EcdispNL})
with respect to the cutoff for the reciprocal lattice vector $G_{\text{max}}$
of the plane-wave expansion of
$\rho_{\text{s}}=\sum_{G<G_{\text{max}}}
\rho_{\text{s}}^{\bm{G}}e^{i\bm{G}\cdot\bm{r}}$
showed that, indeed, $n=1$ requires a smaller $G_{\text{max}}$ than
larger $n$. Therefore, for efficiency $n=1$ is a better choice,
but the price to pay is to have second and higher derivatives of
$\rho_{\text{s}}$ that are discontinuous, which is not elegant but
of no consequence from the practical point of view according to
our tests. $n=1$ in Eq.~(\ref{rhos}) has been chosen for all calculations
presented in Sec.~\ref{results}.

Figure \ref{fig_rhos_2} shows $\rho_{\text{s}}$ for different values of
$\rho_{\text{c}}$ (with $n=1$). For Li for instance, the cutoffs
$\rho_{\text{c}}=0.5$, 1.0, and 1.5~Bohr$^{-3}$ lead to a change in the
density in the region corresponding to a distance $d$ from the nucleus
that is smaller than approximately 0.55, 0.45, and 0.35~Bohr, respectively.
The crucial point is to choose $\rho_{\text{c}}$ such that
a good balance between computational efficiency and accuracy is achieved.
This will be studied in detail in Sec.~\ref{results}.

Like the all-electron density $\rho$, $\rho_{\text{s}}$
has a kink at the nucleus, however, if $\rho$ at the nucleus is much larger
than $\rho_{\text{c}}$, then this kink is strongly attenuated
[see Eqs.~(\ref{drhos}) and (\ref{drhosdrho})].
On a scale like in Figs.~\ref{fig_rhos_1} and \ref{fig_rhos_2},
a kink is clearly visible only for the very lightest atoms.

The RPS method is described in detail in Ref.~\onlinecite{RomanPerezPRL09},
thus only the basic idea is now briefly summarized. The kernel $\Phi$ in
Eq.~(\ref{EcdispNL}) depends on $\rho$ and $\nabla\rho$ at $\bm{r}_{1}$ and
$\bm{r}_{2}$ via a function
$q_{0}\left(\rho,\left\vert\nabla\rho\right\vert\right)$
that is evaluated at $\bm{r}_{1}$ and $\bm{r}_{2}$.
Such a dependency of $\Phi$ on $\bm{r}_{1}$ and $\bm{r}_{2}$
individually, and not only on $r_{12}=\left\vert\bm{r}_{1}-\bm{r}_{2}\right\vert$,
prevents Eq.~(\ref{EcdispNL}) to be evaluated by a single convolution.
Therefore, Rom\'{a}n-P\'{e}rez and Soler proposed to expand $\Phi$
with an interpolation formula consisting of factorized terms,
\begin{equation}
\Phi\left(q_{1},q_{2},r_{12}\right) = \sum_{\alpha,\beta}
\Phi\left(q_{\alpha},q_{\beta},r_{12}\right)
p_{\alpha}(q_{1})p_{\beta}(q_{2}),
\label{phiq1q2}
\end{equation}
allowing the use of convolution that is performed efficiently with FFT.
The sum on the right-hand side of Eq.~(\ref{phiq1q2}) runs over a
two-dimensional mesh $(q_{\alpha},q_{\beta})$ of values of $q_{0}$ at which
$\Phi$ is pre-calculated and multiplied by cubic polynomials $p_{\alpha}$
evaluated at
$q_{i}=q_{0}\left(\rho(\bm{r}_{i}),\left\vert\nabla\rho(\bm{r}_{i})\right\vert\right)$
($i=1, 2$). The accuracy of the interpolation is
determined by the number of points $N_{q}$ on the $q$-mesh, and
the chosen cutoff value $q_{0}^{\text{c}}$ should be larger than all values
of $q_{0}$ in the unit cell. Not explicitly shown in Eq.~(\ref{phiq1q2}),
an interpolation over $r_{12}$ [or equivalently over $k$ for the Fourier
transform $\Phi\left(q_{\alpha},q_{\beta},k\right)]$ is also done.
A careful study of the influence
of $N_{q}$ and $q_{0}^{\text{c}}$ on the lattice constant
and cohesive energy of solids was reported in Ref.~\onlinecite{KlimesPRB11}.
It was concluded that $N_{q}=30$ and $q_{0}^{\text{c}}=10$ lead to
results that are well converged, therefore these parameters were chosen
for our calculations. However, in two cases, namely K and Cs,
we observed that a rather large change in the lattice constant
(of the order of 0.02~\AA) was obtained by increasing
$N_{q}$ to 40. Thus, the results in Sec.~\ref{results} for K and Cs
were obtained with $N_{q}=40$ and $q_{0}^{\text{c}}=10$, while
for all other solids $N_{q}=30$ was used.
We mention that Wu and Gygi\cite{WuJCP12} pointed out that
$q_{1}q_{2}\Phi$ is smoother than $\Phi$, such that it is
computationally more advantageous to expand
the former instead of the latter, since the results converge faster with
respect to $N_{q}$. In Ref.~\onlinecite{CorsettiJCP13}, a similar idea was
applied to the kernel VV10 of Vydrov and Van Voorhis.\cite{VydrovJCP10}

The kernel $\Phi$ that we have considered for the present work is
DRSLL.\cite{DionPRL04}
Extensions of Eq.~(\ref{EcdispNL}) for spin-polarized systems
were proposed in Refs.~\onlinecite{ObataJPSJ13,ThonhauserPRL15}.
However, since our results will be compared to the results from
Refs.~\onlinecite{KlimesPRB11,SchimkaPRB13,ParkCAP15}, we followed
the procedure in these studies which consists of simply using the
sum of the up and down electron densities to evaluate
Eq.~(\ref{EcdispNL}).\cite{KlimesPRB11} This concerns
the calculations for solid Fe and Ni, and most free atoms.

Concerning the potential
$v_{\text{c,disp}}^{\text{NL}}=\delta E_{\text{c,disp}}^{\text{NL}}/\delta\rho$,
it was shown in Ref.~\onlinecite{ThonhauserPRB07} that adding
$v_{\text{c,disp}}^{\text{NL}}$
to the semilocal component of the exchange-correlation potential for
self-consistent calculations leads
to no visible change in the total energy curve.
In other words, the effect of $v_{\text{c,disp}}^{\text{NL}}$ on
the orbitals and electron density is too small to affect properties
calculated with the total energy.
However, for the geometry optimization
with the forces and stress tensor,\cite{SabatiniJPCM12}
the complete potential is required, therefore it is still desirable
to have access to $v_{\text{c,disp}}^{\text{NL}}$.
As explained in more detail in Appendix \ref{appendixB}, the proper calculation
of $v_{\text{c,disp}}^{\text{NL}}$ requires the calculation of
$d\rho_{\text{s}}/d\rho$, which is trivially done with
Eq.~(\ref{rhos}) of the present scheme.

The LAPW calculations were done with the WIEN2k code\cite{WIEN2k} and
the parameters of the calculations like the basis-set size or number of
$\bm{k}$-points for Brillouin zone integrations were chosen to be very well
converged. For the RPS method, the subroutines available in the
QUANTUM ESPRESSO code\cite{GiannozziJPCM09,SabatiniJPCM12} were used and
modified. The FFT were done using version 3.3.5 of the FFTW software
package,\cite{FFTW} which is efficiently parallelized with MPI.

\section{\label{results}Results and discussion}

\begin{figure}
\includegraphics[width=\columnwidth]{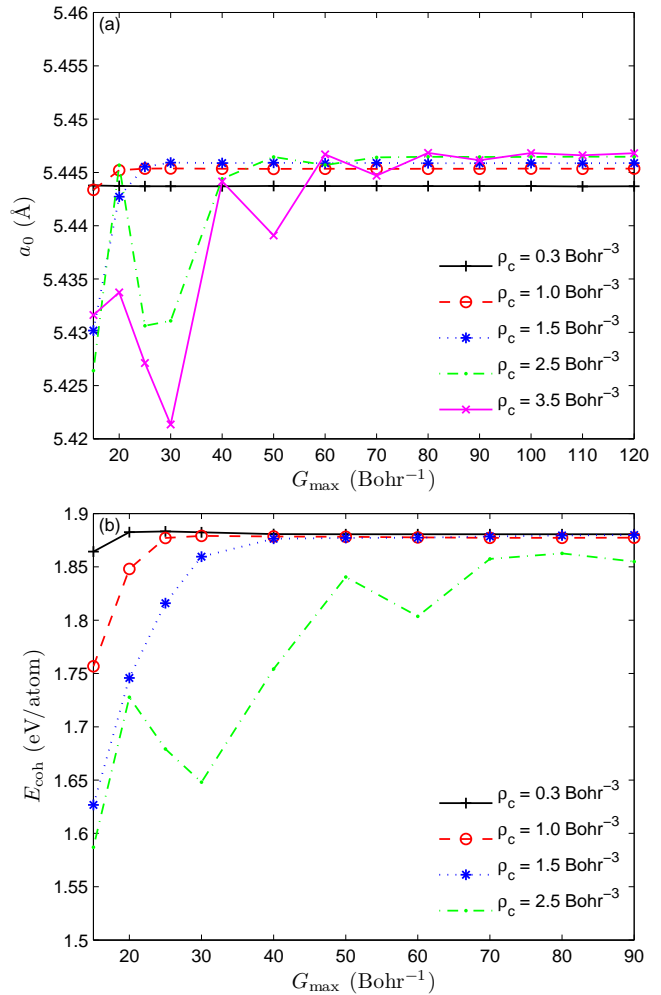}
\caption{\label{fig_Ca}Convergence of the optB88-vdW equilibrium lattice constant $a_{0}$
(a) and cohesive energy $E_{\text{coh}}$ (b) of Ca with respect to the density cutoff
$\rho_{\text{c}}$ and plane-wave expansion cutoff $G_{\text{max}}$.}
\end{figure}
\begin{figure}
\includegraphics[width=\columnwidth]{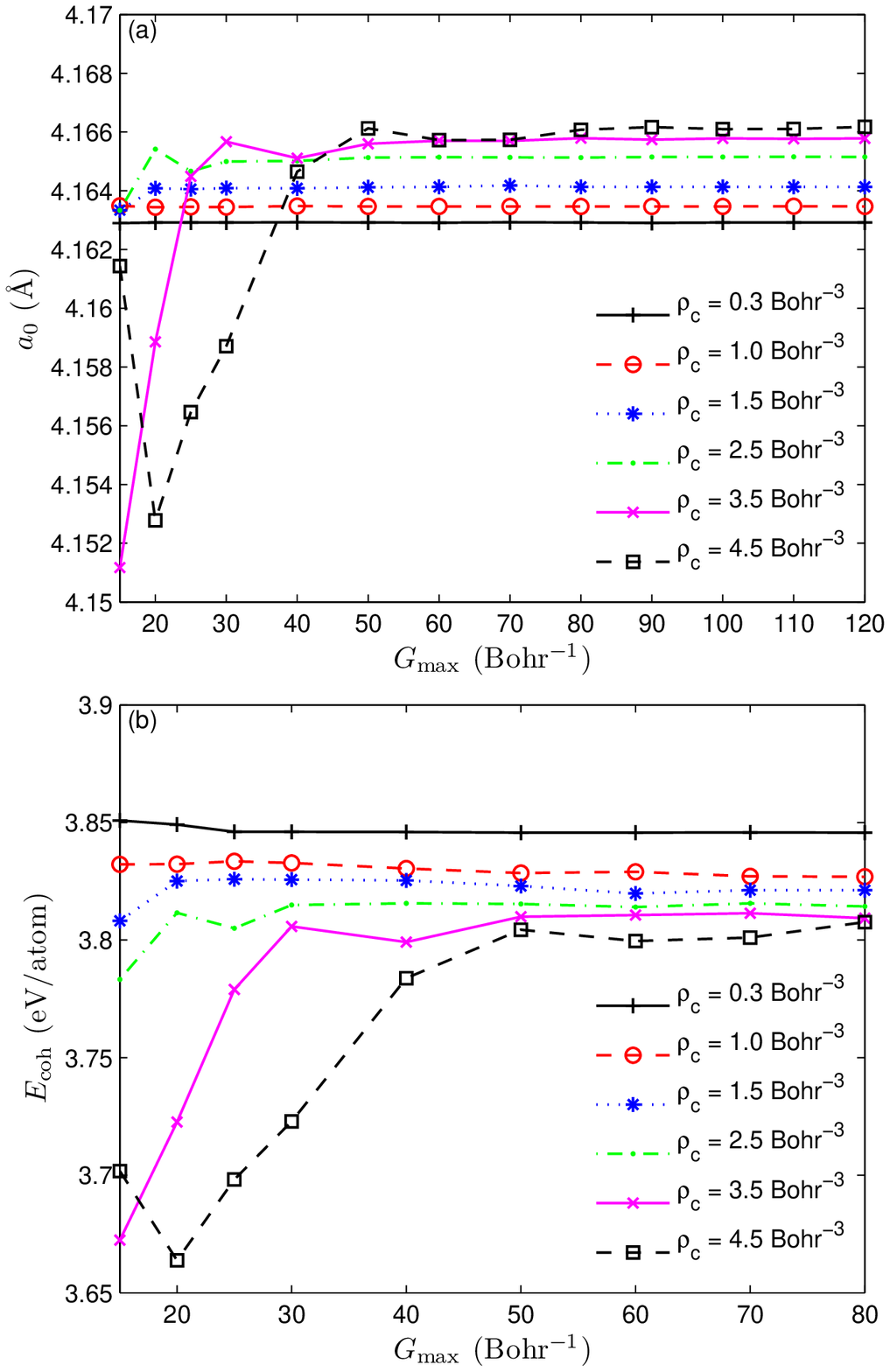}
\caption{\label{fig_Au}Convergence of the optB88-vdW equilibrium lattice constant $a_{0}$
(a) and cohesive energy $E_{\text{coh}}$ (b) of Au with respect to the density cutoff
$\rho_{\text{c}}$ and plane-wave expansion cutoff $G_{\text{max}}$.}
\end{figure}
\begin{figure}
\includegraphics[width=\columnwidth]{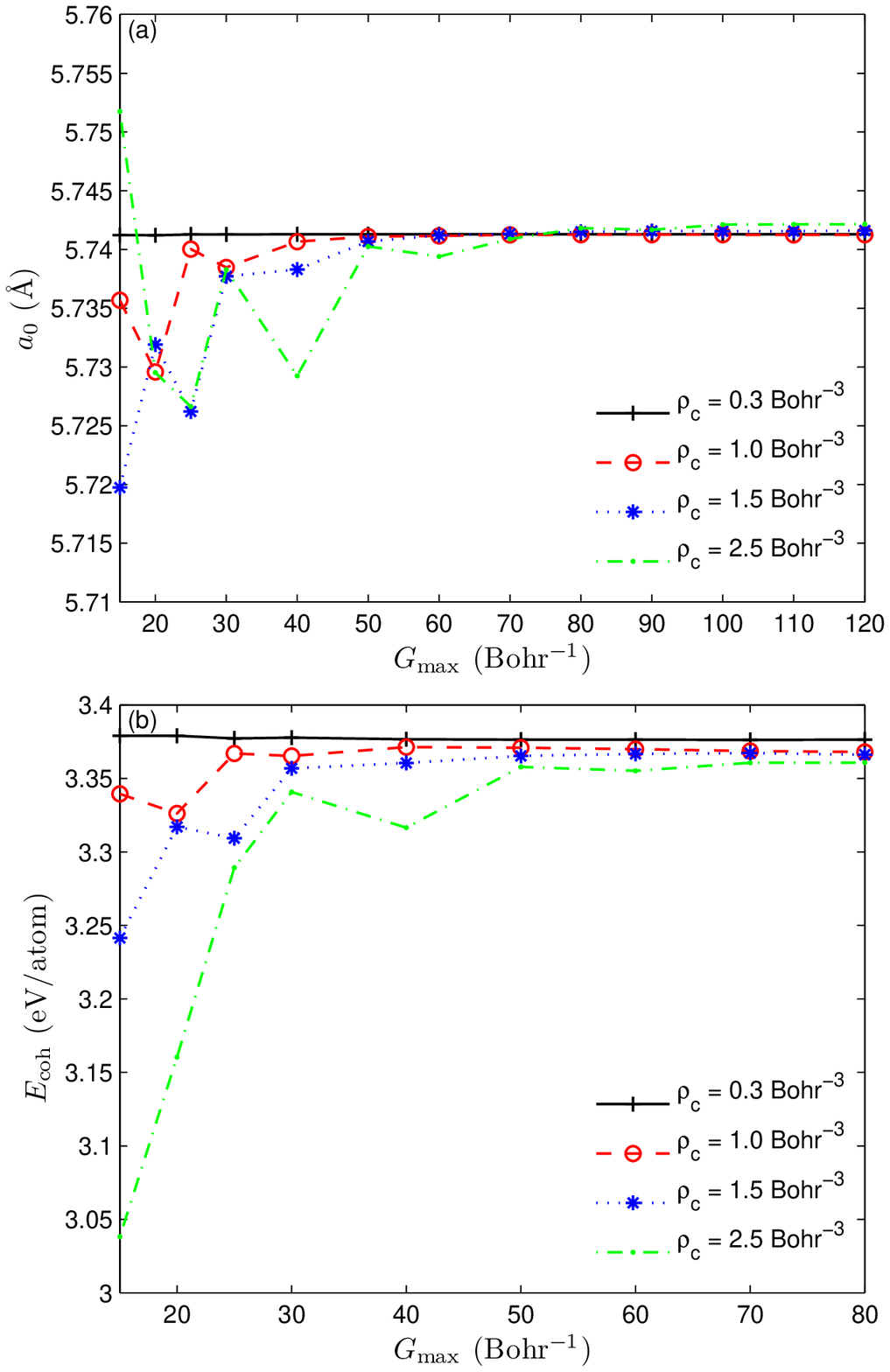}
\caption{\label{fig_GaAs}Convergence of the optB88-vdW equilibrium lattice constant $a_{0}$
(a) and cohesive energy $E_{\text{coh}}$ (b) of GaAs with respect to the density cutoff
$\rho_{\text{c}}$ and plane-wave expansion cutoff $G_{\text{max}}$.}
\end{figure}
\begin{figure}
\includegraphics[width=\columnwidth]{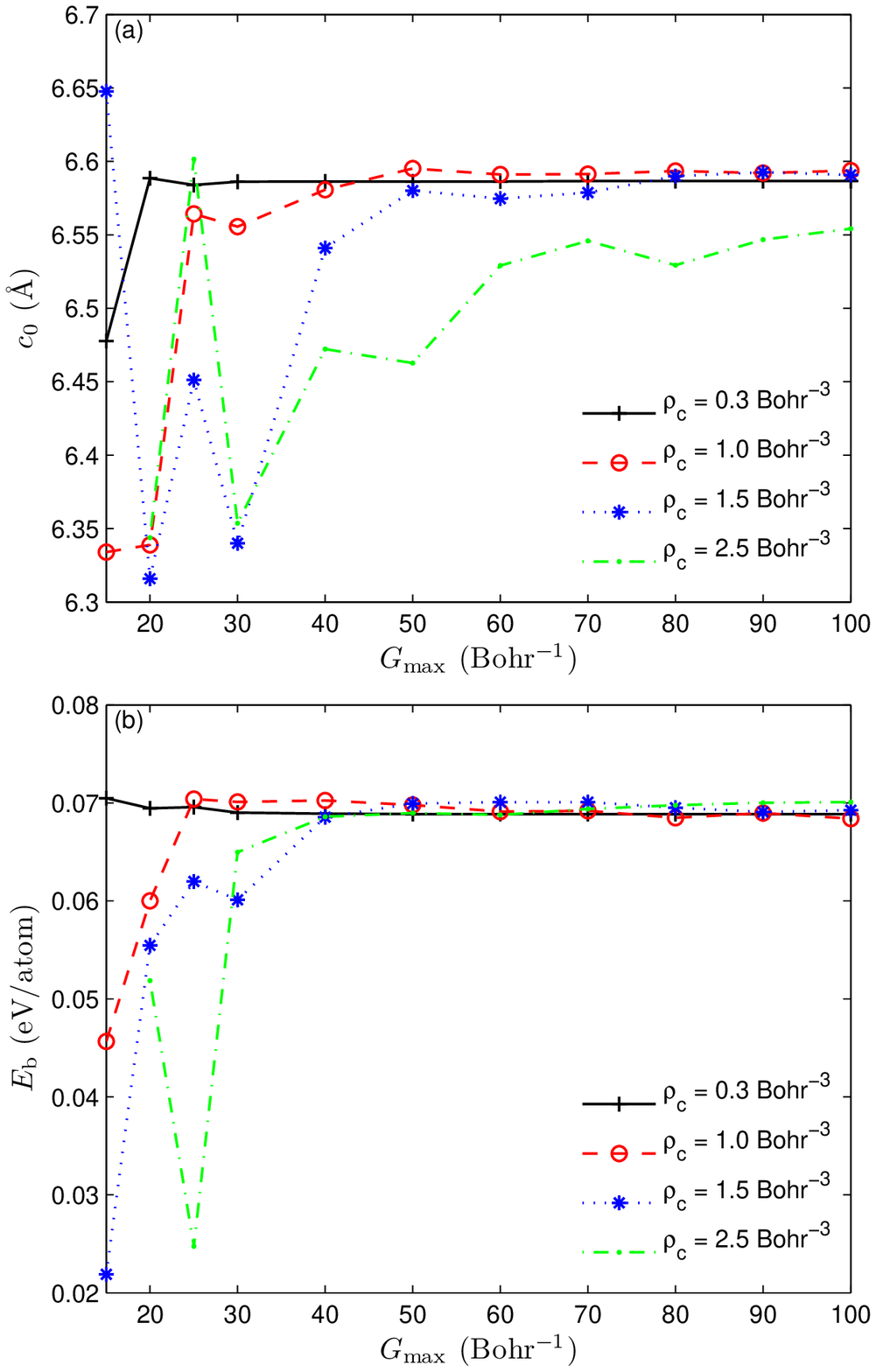}
\caption{\label{fig_hBN}Convergence of the optB88-vdW equilibrium lattice constant $c_{0}$
(a) and interlayer binding energy $E_{\text{b}}$ (b) of h-BN with respect to the density
cutoff $\rho_{\text{c}}$ and plane-wave expansion cutoff $G_{\text{max}}$.}
\end{figure}

The objective of this section is twofold. First,
we will discuss the convergence behavior of the lattice
constant and binding energy with respect to the density cutoff $\rho_{\text{c}}$
and plane-wave expansion cutoff $G_{\text{max}}$.
Details will be shown for a few representative solids.
Then, for a large set of solids (listed in Tables~\ref{aEc}-\ref{LC}),
we will compare our results with the values that were obtained with
other implementations of the nonlocal vdW functionals.
All results presented in this section were obtained with the functional
optB88-vdW,\cite{KlimesJPCM10,KlimesPRB11} whose semilocal component
consists of a modification of the GGA B88\cite{BeckePRA88} for exchange (optB88) and
LDA\cite{PerdewPRB92a}
for correlation, while the nonlocal term Eq.~(\ref{EcdispNL})
uses the DRSLL\cite{DionPRL04} kernel $\Phi$.
optB88-vdW is one of the functionals proposed in
Refs.~\onlinecite{KlimesJPCM10,KlimesPRB11} that have been shown to be
more accurate for both molecules and solids than the older functionals
vdW-DF1\cite{DionPRL04} and vdW-DF2.\cite{LeePRB10}

\subsection{\label{convergence}Convergence tests}

As described in Sec.~\ref{methodology}, the cutoff $\rho_{\text{c}}$ in
Eq.~(\ref{rhos}), which determines the degree of smoothness applied to $\rho$, has
to be chosen. This is a tradeoff between computational efficiency
and accuracy, since a smaller $\rho_{\text{c}}$ allows for a smaller $G_{\text{max}}$
(see below), but increases the deviation of the smooth
density $\rho_{\text{s}}$ from the true density $\rho$ and, therefore, also the value
of the vdW energy [Eq.~(\ref{EcdispNL})].

Starting the discussion with the convergence with respect to the
density cutoff $\rho_{\text{c}}$, we have observed that for all cases
in Tables~\ref{aEc}-\ref{LC} except Ne,
a cutoff $\rho_{\text{c}}=0.3~$Bohr$^{-3}$ provides results for the geometry
and binding energy that are already quite well converged. Compared to
fully converged results (that are obtained with $\rho_{\text{c}}$
between 1~Bohr$^{-3}$ and 3~Bohr$^{-3}$ depending on the solid), the
difference is below $\sim0.005$~\AA~for the lattice constant
and 0.04~eV/atom for the binding energy. The only exception is
the very weakly bound rare gas Ne, since the converged lattice
constant is about 0.02~\AA~larger than the value obtained
with $\rho_{\text{c}}=0.3$~Bohr$^{-3}$.
Nevertheless, for vdW systems, an uncertainty in the lattice
constant or bond length of the order of a few 0.01~\AA~is usually acceptable.

In the examples shown in Figs.~\ref{fig_Ca}-\ref{fig_GaAs} for the cubic
solids Ca, Au, and GaAs,
the lattice constants $a_{0}$ calculated with $\rho_{\text{c}}=0.3$~Bohr$^{-3}$
differ from the converged values
by 0.003, 0.002, and 0.001~\AA, respectively.
For the cohesive energy $E_{\text{coh}}$, the error is the largest
for Au (0.03~eV/atom). Also in the case of hexagonal BN (Fig.~\ref{fig_hBN}), the
results for the lattice constant $c_{0}$ (the interlayer distance is $c_{0}/2$)
and interlayer binding energy $E_{\text{b}}$
are very well converged with $\rho_{\text{c}}=0.3$~Bohr$^{-3}$.

Thus, these results show that
$\rho_{\text{c}}=0.3$~Bohr$^{-3}$ is a pretty good choice in terms
of accuracy for all solids, including vdW systems. This shows that
all-electron benchmark results can be obtained even with a
smooth density $\rho_{\text{s}}$ which differs considerably from
the true density in the region close to the nuclei
(see Figs.~\ref{fig_rhos_1}-\ref{fig_rhos_2}).

At that point we should mention that Ref.~\onlinecite{KlimesPRB11} reports a
comparison between various schemes for the calculation of nonlocal
vdW functionals. The goal was to estimate the accuracy of the PAW
method (the density consists of a sum of pseudovalence and
soft-core components) compared to all-electron results.
A smoothing procedure is also used, however, no details are given,
and also very little is mentioned about the procedure to get the all-electron results.
The conclusion from this study is that the PAW results for the lattice constant
and cohesive energy show good agreement with the all-electron results, similar
to our conclusion, as shown below. However, note that no van der Waals systems
were considered in Ref.~\onlinecite{KlimesPRB11}.

Concerning the convergence with the plane-wave cutoff $G_{\text{max}}$,
we have observed that for most solids, $G_{\text{max}}=15$~Bohr$^{-1}$
is enough for the density cutoff $\rho_{\text{c}}=0.3$~Bohr$^{-3}$.
This is what is shown in Figs.~\ref{fig_Ca}-\ref{fig_GaAs} for
Ca, Au, and GaAs.
With $G_{\text{max}}=15$~Bohr$^{-1}$ the lattice constant and cohesive
energy are usually converged within a few 0.001~\AA~and 0.01~eV/atom, respectively.
However, in the case of weakly bound systems like h-BN (Fig.~\ref{fig_hBN}),
a larger $G_{\text{max}}$ of 20-25~Bohr$^{-1}$ should be used
for well converged results. In order to be on the safe side,
a $G_{\text{max}}$ of about 20~Bohr$^{-1}$ is also recommended for the alkali
metals, which are rather soft systems with a core-core vdW attraction
that should not be neglected.\cite{TaoPRB10}
As clearly shown in Figs.~\ref{fig_Ca}-\ref{fig_hBN}, the
larger $\rho_{\text{c}}$ is, the larger $G_{\text{max}}$ should be in order to
properly expand the density $\rho_{\text{s}}$.

From these convergence tests, it can be concluded that smoothing the
all-electron density with a cutoff $\rho_{\text{c}}=0.3$~Bohr$^{-3}$ is quite
safe in terms of accuracy. Furthermore, this value of $\rho_{\text{c}}$
is appropriate for the very light as well as for the very heavy atoms.
With $\rho_{\text{c}}=0.3$~Bohr$^{-3}$, a density
plane-wave expansion in the range 15$-$20~Bohr$^{-1}$ has to be used
(25~Bohr$^{-1}$ for the very weakly bound Ne), which seems to be similar
to what may be needed to get converged results for weakly bound systems
with pseudopotentials methods
(see Ref.~\onlinecite{SorescuJCTC14}).

\subsection{\label{comparison}Comparison with other codes}

\begin{table*}
\caption{\label{aEc}Equilibrium lattice constant $a_{0}$ (in \AA) and
cohesive energy $E_{\text{coh}}$ (in eV/atom) of 35 cubic solids.
The space group number is indicated in parenthesis. An estimate of the error bar for
the WIEN2k results is 0.003~\AA~for $a_{0}$ and
0.03~eV/atom for $E_{\text{coh}}$.}
\begin{ruledtabular}
\begin{tabular}{lcccccccc}
\multicolumn{1}{l}{} &
\multicolumn{4}{c}{$a_{0}$} &
\multicolumn{4}{c}{$E_{\text{coh}}$} \\
\cline{2-5}\cline{6-9}
\multicolumn{1}{l}{} &
\multicolumn{2}{c}{PBE} &
\multicolumn{2}{c}{optB88-vdW} &
\multicolumn{2}{c}{PBE} &
\multicolumn{2}{c}{optB88-vdW} \\
\cline{2-3}\cline{4-5}\cline{6-7}\cline{8-9}
\multicolumn{1}{l}{Solid} &
\multicolumn{1}{c}{WIEN2k} &
\multicolumn{1}{c}{VASP} &
\multicolumn{1}{c}{WIEN2k} &
\multicolumn{1}{c}{VASP} &
\multicolumn{1}{c}{WIEN2k} &
\multicolumn{1}{c}{VASP} &
\multicolumn{1}{c}{WIEN2k} &
\multicolumn{1}{c}{VASP} \\
\hline
Li (229)   & 3.437 &3.437,\footnotemark[1]			3.439\footnotemark[3] & 3.433 &3.432,\footnotemark[1]			   3.435\footnotemark[3] & 1.61 &1.60,\footnotemark[1]  		   1.61\footnotemark[3] & 1.58& 1.57,\footnotemark[1]1.59\footnotemark[3] \\
Na (229)   & 4.200 &4.200,\footnotemark[1]			4.193\footnotemark[3] & 4.153 &4.169,\footnotemark[1]			   4.152\footnotemark[3] & 1.08 &1.08,\footnotemark[1]  		   1.09\footnotemark[3] & 1.05& 1.04,\footnotemark[1]1.07\footnotemark[3] \\
K (229)    & 5.283 &5.284,\footnotemark[1]5.278,\footnotemark[2]5.284\footnotemark[3] & 5.170 &5.168,\footnotemark[1]5.159,\footnotemark[2]5.162\footnotemark[3] & 0.87 &0.86,\footnotemark[1]0.87,\footnotemark[2]0.87\footnotemark[3] & 0.88& 0.88,\footnotemark[1]0.89\footnotemark[3] \\
Rb (229)   & 5.671 &5.671,\footnotemark[1]5.667,\footnotemark[2]5.668\footnotemark[3] & 5.506 &5.506,\footnotemark[1]5.535,\footnotemark[2]5.501\footnotemark[3] & 0.78 &0.77,\footnotemark[1]0.77,\footnotemark[2]0.77\footnotemark[3] & 0.82& 0.81,\footnotemark[1]0.83\footnotemark[3] \\
Cs (229)   & 6.162 &6.160,\footnotemark[1]6.156,\footnotemark[2]6.161\footnotemark[3] & 5.915 &5.899,\footnotemark[1]5.900,\footnotemark[2]5.899\footnotemark[3] & 0.72 &0.70,\footnotemark[1]0.72,\footnotemark[2]0.72\footnotemark[3] & 0.79& 0.79,\footnotemark[1]0.66\footnotemark[3] \\
Ca (225)   & 5.528 &5.533,\footnotemark[1]5.524,\footnotemark[2]5.532\footnotemark[3] & 5.446 &5.450,\footnotemark[1]5.443,\footnotemark[2]5.450\footnotemark[3] & 1.91 &1.90,\footnotemark[1]1.91,\footnotemark[2]1.90\footnotemark[3] & 1.88& 1.88,\footnotemark[1]1.88\footnotemark[3] \\
Sr (225)   & 6.025 &6.019,\footnotemark[1]6.020,\footnotemark[2]6.026\footnotemark[3] & 5.919 &5.917,\footnotemark[1]5.917,\footnotemark[2]5.911\footnotemark[3] & 1.61 &1.61,\footnotemark[1]1.61,\footnotemark[2]1.61\footnotemark[3] & 1.60& 1.61,\footnotemark[1]1.61\footnotemark[3] \\
Ba (229)   & 5.020 &5.028,\footnotemark[1]5.018,\footnotemark[2]5.030\footnotemark[3] & 4.905 &4.917,\footnotemark[1]4.903,\footnotemark[2]4.915\footnotemark[3] & 1.89 &1.88,\footnotemark[1]1.88,\footnotemark[2]1.88\footnotemark[3] & 1.99& 1.99,\footnotemark[1]2.00\footnotemark[3] \\
Al (225)   & 4.041 &4.041,\footnotemark[1]			4.040\footnotemark[3] & 4.054 &4.054,\footnotemark[1]			   4.054\footnotemark[3] & 3.44 &3.50,\footnotemark[1]  		   3.54\footnotemark[3] & 3.24& 3.34,\footnotemark[1]3.38\footnotemark[3] \\
C (227)    & 3.575 &3.574,\footnotemark[1]			3.573\footnotemark[3] & 3.577 &3.577,\footnotemark[1]			   3.575\footnotemark[3] & 7.71 &7.70,\footnotemark[1]  		   7.85\footnotemark[3] & 7.72& 7.70,\footnotemark[1]7.89\footnotemark[3] \\
Si (227)   & 5.471 &5.465,\footnotemark[1]			5.469\footnotemark[3] & 5.464 &5.460,\footnotemark[1]			   5.469\footnotemark[3] & 4.57 &4.62,\footnotemark[1]  		   4.61\footnotemark[3] & 4.67& 4.74,\footnotemark[1]4.74\footnotemark[3] \\
Ge (227)   & 5.764 &5.766,\footnotemark[1]			5.783\footnotemark[3] & 5.762 &5.762,\footnotemark[1]			   5.798\footnotemark[3] & 3.73 &3.72,\footnotemark[1]  		   3.74\footnotemark[3] & 3.93& 3.90,\footnotemark[1]3.92\footnotemark[3] \\
Sn (227)   & 6.657 &						6.652\footnotemark[3] & 6.640 & 					   6.637\footnotemark[3] & 3.17 &					   3.20\footnotemark[3] & 3.42& 		     3.44\footnotemark[3] \\
Pb (225)   & 5.034 &						5.031\footnotemark[3] & 5.026 & 					   5.018\footnotemark[3] & 2.98 &					   2.98\footnotemark[3] & 3.32& 		     3.32\footnotemark[3] \\
V (229)    & 2.998 &			  2.996,\footnotemark[2]2.978\footnotemark[3] & 2.986 & 		     2.984,\footnotemark[2]2.969\footnotemark[3] & 5.37 &		      5.37,\footnotemark[2]5.41\footnotemark[3] & 5.63& 		     5.74\footnotemark[3] \\
Fe (229)   & 2.832 &			  2.832,\footnotemark[2]2.832\footnotemark[3] & 2.820 & 		     2.821,\footnotemark[2]2.821\footnotemark[3] & 4.92 &		      4.89,\footnotemark[2]5.16\footnotemark[3] & 5.03& 		     5.11\footnotemark[3] \\
Ni (225)   & 3.518 &			  3.507,\footnotemark[2]3.518\footnotemark[3] & 3.513 & 		     3.512,\footnotemark[2]3.511\footnotemark[3] & 4.76 &		      4.75,\footnotemark[2]4.80\footnotemark[3] & 4.85& 		     4.98\footnotemark[3] \\
Cu (225)   & 3.632 &3.635,\footnotemark[1]3.631,\footnotemark[2]3.635\footnotemark[3] & 3.630 &3.632,\footnotemark[1]3.622,\footnotemark[2]3.629\footnotemark[3] & 3.52 &3.49,\footnotemark[1]3.50,\footnotemark[2]3.49\footnotemark[3] & 3.59& 3.52,\footnotemark[1]3.57\footnotemark[3] \\
Nb (229)   & 3.312 &			  3.310,\footnotemark[2]3.308\footnotemark[3] & 3.306 & 		     3.307,\footnotemark[2]3.303\footnotemark[3] & 6.98 &		      6.96,\footnotemark[2]7.06\footnotemark[3] & 7.42& 		     7.49\footnotemark[3] \\
Mo (229)   & 3.162 &			  3.161,\footnotemark[2]3.151\footnotemark[3] & 3.162 & 		     3.162,\footnotemark[2]3.154\footnotemark[3] & 6.28 &		      6.28,\footnotemark[2]7.76\footnotemark[3] & 6.85& 		     6.95\footnotemark[3] \\
Rh (225)   & 3.832 &3.830,\footnotemark[1]3.831,\footnotemark[2]3.824\footnotemark[3] & 3.835 &3.831,\footnotemark[1]3.834,\footnotemark[2]3.829\footnotemark[3] & 5.74 &5.82,\footnotemark[1]5.70,\footnotemark[2]6.02\footnotemark[3] & 6.06& 6.10,\footnotemark[1]6.34\footnotemark[3] \\
Pd (225)   & 3.943 &3.943,\footnotemark[1]3.933,\footnotemark[2]3.942\footnotemark[3] & 3.940 &3.941,\footnotemark[1]3.930,\footnotemark[2]3.938\footnotemark[3] & 3.71 &3.71,\footnotemark[1]3.76,\footnotemark[2]3.74\footnotemark[3] & 4.03& 3.96,\footnotemark[1]4.04\footnotemark[3] \\
Ag (225)   & 4.148 &4.154,\footnotemark[1]4.145,\footnotemark[2]4.147\footnotemark[3] & 4.136 &4.141,\footnotemark[1]4.127,\footnotemark[2]4.130\footnotemark[3] & 2.53 &2.50,\footnotemark[1]2.52,\footnotemark[2]2.52\footnotemark[3] & 2.81& 2.76,\footnotemark[1]2.82\footnotemark[3] \\
Ta (229)   & 3.320 &			  3.317,\footnotemark[2]3.309\footnotemark[3] & 3.311 & 		     3.307,\footnotemark[2]3.306\footnotemark[3] & 8.22 &		      8.11,\footnotemark[2]8.41\footnotemark[3] & 8.33& 		     8.50\footnotemark[3] \\
W (229)    & 3.185 &			  3.182,\footnotemark[2]3.172\footnotemark[3] & 3.184 & 		     3.181,\footnotemark[2]3.178\footnotemark[3] & 8.34 &		      8.39,\footnotemark[2]8.48\footnotemark[3] & 8.84& 		     9.01\footnotemark[3] \\
Ir (225)   & 3.874 &			  3.868,\footnotemark[2]3.873\footnotemark[3] & 3.882 & 		     3.879,\footnotemark[2]3.886\footnotemark[3] & 7.35 &		      7.31,\footnotemark[2]7.28\footnotemark[3] & 7.63& 		     7.60\footnotemark[3] \\
Pt (225)   & 3.971 &			  3.969,\footnotemark[2]3.968\footnotemark[3] & 3.979 & 		     3.978,\footnotemark[2]3.980\footnotemark[3] & 5.55 &		      5.51,\footnotemark[2]5.58\footnotemark[3] & 5.85& 		     5.90\footnotemark[3] \\
Au (225)   & 4.161 &			  4.154,\footnotemark[2]4.157\footnotemark[3] & 4.166 & 		     4.156,\footnotemark[2]4.161\footnotemark[3] & 3.03 &		      3.05,\footnotemark[2]3.04\footnotemark[3] & 3.82& 		     3.40\footnotemark[3] \\
LiF (225)  & 4.070 &4.068\footnotemark[1]					      & 4.032 &4.033\footnotemark[1]						 & 4.33 &4.32\footnotemark[1]					        & 4.54& 4.53\footnotemark[1]			  \\
LiCl (225) & 5.152 &5.152\footnotemark[1]					      & 5.113 &5.114\footnotemark[1]						 & 3.37 &3.42\footnotemark[1]					        & 3.58& 3.61\footnotemark[1]			  \\
NaF (225)  & 4.706 &4.708\footnotemark[1]					      & 4.642 &4.647\footnotemark[1]						 & 3.84 &3.82\footnotemark[1]					        & 4.05& 4.02\footnotemark[1]			  \\
NaCl (225) & 5.700 &5.701\footnotemark[1]					      & 5.617 &5.622\footnotemark[1]						 & 3.10 &3.15\footnotemark[1]					        & 3.31& 3.33\footnotemark[1]			  \\
MgO (225)  & 4.259 &4.257\footnotemark[1]					      & 4.234 &4.231\footnotemark[1]						 & 4.99 &4.97\footnotemark[1]					        & 5.24& 5.21\footnotemark[1]			  \\
SiC (216)  & 4.385 &4.377\footnotemark[1]					      & 4.380 &4.375\footnotemark[1]						 & 6.40 &6.44\footnotemark[1]					        & 6.49& 6.52\footnotemark[1]			  \\
GaAs (216) & 5.749 &5.752\footnotemark[1]					      & 5.742 &5.751\footnotemark[1]						 & 3.15 &3.15\footnotemark[1]					        & 3.37& 3.36\footnotemark[1]			  \\
\end{tabular}
\end{ruledtabular}
\footnotetext[1]{Ref.~\onlinecite{KlimesPRB11}.}
\footnotetext[2]{Ref.~\onlinecite{SchimkaPRB13}.}
\footnotetext[3]{Ref.~\onlinecite{ParkCAP15}.}
\end{table*}

\begin{table*}
\caption{\label{RG}Equilibrium lattice constant $a_{0}$ (in \AA) and
cohesive energy $E_{\text{coh}}$ (in meV/atom) of rare-gas solids.
The space group number is indicated in parenthesis. An estimate of the
error bar for the WIEN2k results is 0.01~\AA~for $a_{0}$ and
5~meV/atom for $E_{\text{coh}}$.}
\begin{ruledtabular}
\begin{tabular}{lcccccccccc}
\multicolumn{1}{l}{} &
\multicolumn{5}{c}{$a_{0}$} &
\multicolumn{5}{c}{$E_{\text{coh}}$} \\
\cline{2-6}\cline{7-11}
\multicolumn{1}{l}{} &
\multicolumn{2}{c}{LDA} &
\multicolumn{3}{c}{optB88-vdW} &
\multicolumn{2}{c}{LDA} &
\multicolumn{3}{c}{optB88-vdW} \\
\cline{2-3}\cline{4-6}\cline{7-8}\cline{9-11}
\multicolumn{1}{l}{Solid} &
\multicolumn{1}{c}{WIEN2k} &
\multicolumn{1}{c}{CP2K\footnotemark[1]} &
\multicolumn{1}{c}{WIEN2k} &
\multicolumn{1}{c}{CP2K\footnotemark[1]} &
\multicolumn{1}{c}{VASP\footnotemark[2]} &
\multicolumn{1}{c}{WIEN2k} &
\multicolumn{1}{c}{CP2K\footnotemark[1]} &
\multicolumn{1}{c}{WIEN2k} &
\multicolumn{1}{c}{CP2K\footnotemark[1]} &
\multicolumn{1}{c}{VASP\footnotemark[2]} \\
\hline
Ne (225) & 3.86 & 3.86 & 4.27 & 4.24 & 4.25 &  87 &  92 &  49 &  59 &  45 \\
Ar (225) & 4.94 & 4.94 & 5.24 & 5.24 & 5.22 & 138 & 136 & 136 & 143 & 143 \\
Kr (225) & 5.33 & 5.36 & 5.63 & 5.63 & 5.61 & 169 & 164 & 179 & 181 & 180 \\
\end{tabular}
\end{ruledtabular}
\footnotetext[1]{Ref.~\onlinecite{TranJCP13}.}
\footnotetext[2]{Ref.~\onlinecite{CallsenPRB15}.}
\end{table*}

\begin{table*}
\caption{\label{LC}Equilibrium lattice constant $c_{0}$ (in \AA) and interlayer
binding energy $E_{\text{b}}$ (in meV/atom) of hexagonal layered solids.
The intralayer lattice constant $a$ was kept fixed at the experimental value
of 2.462 and 2.503~\AA~for graphite and h-BN, respectively. The space group
number is indicated in parenthesis. An estimate of the error bar for the WIEN2k
results is 0.03~\AA~for $c_{0}$ and 5~meV/atom for $E_{\text{b}}$.}
\begin{ruledtabular}
\begin{tabular}{lcccccccc}
\multicolumn{1}{l}{} &
\multicolumn{4}{c}{$c_{0}$} &
\multicolumn{4}{c}{$E_{\text{b}}$} \\
\cline{2-5}\cline{6-9}
\multicolumn{1}{l}{} &
\multicolumn{2}{c}{LDA} &
\multicolumn{2}{c}{optB88-vdW} &
\multicolumn{2}{c}{LDA} &
\multicolumn{2}{c}{optB88-vdW} \\
\cline{2-3}\cline{4-5}\cline{6-7}\cline{8-9}
\multicolumn{1}{l}{Solid} &
\multicolumn{1}{c}{WIEN2k} &
\multicolumn{1}{c}{VASP} &
\multicolumn{1}{c}{WIEN2k} &
\multicolumn{1}{c}{VASP} &
\multicolumn{1}{c}{WIEN2k} &
\multicolumn{1}{c}{VASP} &
\multicolumn{1}{c}{WIEN2k} &
\multicolumn{1}{c}{VASP} \\
\hline
Graphite (194) & 6.68 & 6.62,\footnotemark[1]6.75\footnotemark[2] & 6.72 & 6.72,\footnotemark[1]6.76\footnotemark[3] & 24 & 24,\footnotemark[1]25\footnotemark[2] & 68 & 65,\footnotemark[1]66\footnotemark[3] \\
h-BN (194)     & 6.48 & 6.42,\footnotemark[1]6.58\footnotemark[2] & 6.59 & 6.60,\footnotemark[1]6.64\footnotemark[3] & 28 & 28,\footnotemark[1]28\footnotemark[2] & 69 & 65,\footnotemark[1]67\footnotemark[3] \\
\end{tabular}
\end{ruledtabular}
\footnotetext[1]{Ref.~\onlinecite{GrazianoJPCM12}.}
\footnotetext[2]{Ref.~\onlinecite{BjorkmanPRL12}.}
\footnotetext[3]{Ref.~\onlinecite{BjorkmanJCP14}.}
\end{table*}

Turning now to the comparison with results from the literature,
\cite{KlimesPRB11,SchimkaPRB13,ParkCAP15,TranJCP13,CallsenPRB15,GrazianoJPCM12,BjorkmanPRL12,BjorkmanJCP14}
Tables~\ref{aEc}-\ref{LC} show our converged (with respect to $\rho_{\text{c}}$
and $G_{\text{max}}$) optB88-vdW results for the
lattice constant and binding energy along with results
obtained with the PAW and Gaussian augmented plane wave (GAPW) methods as implemented
into the VASP\cite{KressePRB96} and CP2K codes,\cite{VandeVondeleCPC05} respectively.
Results obtained with the PBE\cite{PerdewPRL96} or LDA functionals
are also shown in order to provide an idea how much (dis)agreement should be
expected between the different codes.

The results in Table~\ref{aEc} for the strongly bound cubic solids show that
the agreement between the WIEN2k and VASP codes is in general excellent.
The differences in the lattice constant $a_{0}$ between the two codes are of
the order of only a few 0.001~\AA~for most solids. We just note that in some cases,
the various VASP results do not agree with each others,
as for instance for Rb and Ge, where the differences are above
0.03~\AA~with optB88-vdW. In such cases, our WIEN2k benchmark results
may help to indicate which one of the VASP results should be more correct.
The WIEN2k and VASP results for the cohesive energy $E_{\text{coh}}$
are overall in quite good agreement as well, since the differences are typically
of the order of a few $0.01$~eV/atom, which is very small.
An exception is Au for which a large
discrepancy is obtained with optB88-vdW (3.82~eV/atom for WIEN2k and 3.40~eV/atom for
VASP\cite{ParkCAP15}), while the PBE results differ by only $\sim0.01$~eV/atom.
Let us also mention that the two VASP results for Mo
with PBE differ considerably (6.28~eV/atom from Ref.~\onlinecite{SchimkaPRB13} and
7.76~eV/atom from Ref.~\onlinecite{ParkCAP15}). Such large disagreements in
$E_{\text{coh}}$, as obtained for Mo and Au, may be due to different electronic
configurations of the $d$-electrons in the isolated atoms.

As a side note, we mention that for the solids in Table~\ref{aEc}, the omission
of the vdW term [Eq.~(\ref{EcdispNL})] in the optB88-vdW functional leads to
lengthening of the lattice constant and reduction of the cohesive energy
that are rather substantial. Our calculations with the functional
$E_{\text{xc}}=E_{\text{x}}^{\text{optB88}}+E_{\text{c}}^{\text{LDA}}$
(results not shown) lead to lattice constants that are larger by
0.05-0.1~\AA~for the transition metals, while for all systems
containing alkali or alkaline earth atoms (except Li and MgO)
$a_{0}$ is larger by 0.1-0.25~\AA. For the cohesive energy, the values
without vdW term are smaller by 0.3-0.6~eV/atom for the systems with alkali and
alkaline earth atoms and 1-1.5~eV/atom for
the transition metals. Therefore, not only vdW systems, but also
those with supposedly unimportant vdW interactions are useful to test the
implementation of functionals specifically designed for vdW systems.
As shown in Ref.~\onlinecite{KlimesPRB11},
optB88-vdW is, compared to PBE, of the same accuracy for the lattice constant
and slightly more accurate for the cohesive energy, such that it can be
considered as rather good for solids. Without the vdW term, the lattice constant
and cohesive energy are compared to experiment largely overestimated and
underestimated, respectively.

Tables~\ref{RG} and \ref{LC} show the results for the weakly bound rare-gas and
hexagonal layered solids, respectively. For vdW systems, it is reasonable to
tolerate uncertainties of a few 0.01~\AA~for the lattice constant and
up to 10 or 20~meV/atom for the binding energy, depending on the system.
From the results it can be inferred
that the agreement between the various codes is rather good and, actually,
on average the discrepancies do not seem to be larger for optB88-vdW than for LDA.
Nevertheless, we note that for Ne, the lattice constant is noticeably larger
(by 0.02-0.03~\AA) with WIEN2k. As noticed above, among all systems
that we have considered,
Ne is the only one for which a density cutoff $\rho_{\text{c}}=0.3$~Bohr$^{-3}$
is not large enough to get a lattice constant that is within 0.01~\AA~of the
converged value. With $\rho_{\text{c}}=0.3$~Bohr$^{-3}$, $a_{0}=4.25$~\AA~which
is closer to the VASP and CP2K values and would possibly indicate that the
PAW and GAPW smooth densities plugged into Eq.~(\ref{EcdispNL}) correspond more
to our density with $\rho_{\text{c}}=0.3$~Bohr$^{-3}$.
In the case of the layered systems graphite and h-BN,
\cite{BjorkmanPRB12,BjorkmanJCP14,RegoJPCM15} the largest
disagreements are for the lattice constant $c_{0}$ calculated with LDA, since the
various VASP values differ by more than 0.15~\AA~which is quite large.
Our LDA results for $c_{0}$ are in between the values from
Refs.~\onlinecite{GrazianoJPCM12,BjorkmanPRL12}.
The optB88-vdW values for $c_{0}$ do not differ that much ($\sim0.05$~\AA), but
the agreement between the codes is also not perfect.
We mention that in our calculations, the
intralayer lattice constant
$a$ was kept fixed at the experimental value of 2.462 and
2.503~\AA~for graphite and h-BN, respectively. The same procedure was used
in Ref.~\onlinecite{BjorkmanPRL12}, while no details are given in
Ref.~\onlinecite{GrazianoJPCM12}.
On the other hand, the interlayer binding energies $E_{\text{b}}$
calculated with WIEN2k and VASP agree quite well.

\subsection{\label{discussion}Further discussion}

\begin{figure}
\includegraphics[width=\columnwidth]{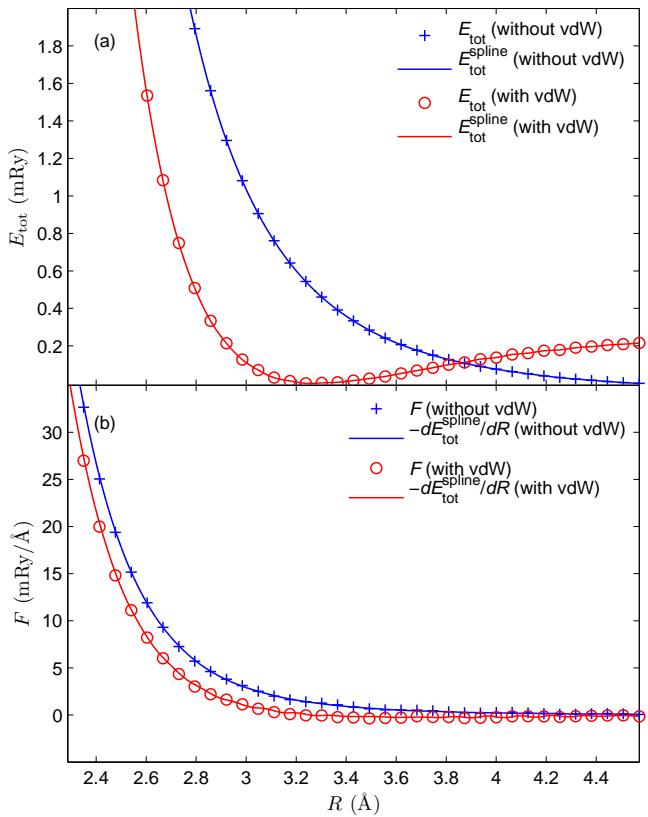}
\caption{\label{fig_Ne2}(a) Total energy (the zero is set at the minimum of the
corresponding curves) of the Ne dimer as a function of the Ne-Ne distance.
The vdW energy was calculated with a smooth density $\rho_{\text{s}}$
corresponding to $\rho_{\text{c}}=0.3$~Bohr$^{-3}$. The solid lines are spline
interpolations of the WIEN2k values represented by the symbols.
(b) Forces acting on the Ne atoms. The solid lines are numerical derivatives
of the spline interpolations of the total energy and the symbols are the values
from WIEN2k.}
\end{figure}

It was mentioned in Sec.~\ref{methodology} that the potential
$v_{\text{c,disp}}^{\text{NL}}=\delta E_{\text{c,disp}}^{\text{NL}}/\delta\rho$
can be calculated (see Appendix \ref{appendixB} for details), which is
necessary for the proper calculation of the forces acting on the nuclei.
The unit cells of the systems that we have considered for our benchmark
calculations do not contain free internal parameters. For such cases
it does not matter how accurately the potential is calculated since it has
basically no effect on the total energy curve.\cite{ThonhauserPRB07}
A way to check the correctness of the implementation of
$v_{\text{c,disp}}^{\text{NL}}$
is to consider the forces in systems with free internal parameters.
We could check that the forces are correct, thus validating our implementation
of the potential. Figure \ref{fig_Ne2} shows the example of the Ne dimer,
where we can see that the calculated forces agree
very well with the numerical derivative of the total energy.
The agreement is excellent in both cases, with or without the vdW term
[Eq.~(\ref{EcdispNL})] in the functional optB88-vdW. According to some tests,
consistency between the total energy and forces is also achieved
with small plane-wave cutoff $G_{\text{max}}$ that maybe not be large enough
for very well converged lattice constant for instance.

As an alternative to the smooth density $\rho_{\text{s}}$ given by
Eq.~(\ref{rhos}), we may use the pseudocharge density generated by
the method of Weinert\cite{WeinertJMP81} for the calculation
of the Coulomb potential in the LAPW method. In this method, the
total (i.e., electronic plus nuclear) charge density inside the atomic
spheres S$_{\alpha}$ is replaced by a pseudodensity $\rho_{\text{s}}^{\text{pseudo}}$ having
the same multipole moments (and therefore producing the same Coulomb potential
in the interstitial), but that is much smoother such that it can be expanded in plane
waves. The only modification that has to be done for the
present purpose is to remove the nuclear contribution. For comparison,
the optB88-vdW lattice constants were also
calculated using the pseudodensity, and the results (not shown) for most
strongly bound solids are very similar to the reference results
in Table~\ref{aEc} obtained with Eq.~(\ref{rhos}).
The largest discrepancies, which are of the order of
0.02~\AA, are found for the alkali metals (except Li) and Ag,
and for these cases the lattice constant is usually larger with
the pseudodensity $\rho_{\text{s}}^{\text{pseudo}}$ than Eq.~(\ref{rhos}).
For the rare-gas solids, the lattice
constants using $\rho_{\text{s}}^{\text{pseudo}}$ are noticeably larger by 0.03 (Ne),
0.05 (Ar), and 0.05~\AA~(Kr), while for graphite and h-BN, $c_{0}$ is larger
with $\rho_{\text{s}}^{\text{pseudo}}$ by 0.01 and 0.03~\AA, respectively, which can be considered
as reasonably small. So, overall the use of the pseudodensity
$\rho_{\text{s}}^{\text{pseudo}}$
leads to results which are quite close to the reference results, except for the
rare gases and a couple of other systems.

Now, let us enumerate the pros and cons of both schemes, Eq.~(\ref{rhos}) and
the pseudocharge method, to generate a smooth density. Equation~(\ref{rhos})
is trivial to implement and basis-set independent, while the pseudocharge
method is a rather complicated method to implement and was designed specifically
for the LAPW method.\cite{WeinertJMP81} An unambiguous way to reach the
converged all-electron results with Eq.~(\ref{rhos}) is to increase the density
cutoff $\rho_{\text{c}}$, while it is not clear how to do it with the
pseudocharge method (however, we admit that we have not checked the influence
of the parameter $n$ in this method\cite{WeinertJMP81}). Finally, the third
advantage with Eq.~(\ref{rhos}) is the possibility to calculate
$d\rho_{\text{s}}/d\rho$ for the potential [see Eq.~(\ref{vcdispNL})],
while it is at present not clear for us how to calculate
$d\rho_{\text{s}}^{\text{pseudo}}/d\rho$
for the pseudodensity. Actually, if
$\rho_{\text{s}}^{\text{pseudo}}\neq\rho_{\text{s}}^{\text{pseudo}}(\rho$),
then this would be impossible even numerically.
On the other hand, the advantage of the pseudocharge method is to lead to
faster calculations, since due to the way $\rho_{\text{s}}^{\text{pseudo}}$
is constructed, it is sufficient to expand it with the plane-wave cutoff $G_{\text{max}}$
for the expansion of the Coulomb potential in the interstitial region
($G_{\text{max}}=12$~Bohr$^{-1}$
is the default in the WIEN2k code). The ideal method should have all advantages
of Eq.~(\ref{rhos}) and produce converged results
with a density plane-wave cutoff that is below 15~Bohr$^{-1}$ for all kinds of systems
including weakly bound systems.

\section{\label{summary}Summary and conclusion}

In summary, a procedure for combining the efficient FFT-based method of RPS
for nonlocal vdW functionals with all-electron methods has been presented and
implemented into an LAPW code. It is based on the simple idea
which consists of smoothing the all-electron density in the core region of the
atoms, and then to use the resulting smooth density in the RPS method.
There are obviously many ways to smooth an all-electron density, but
the one that we have proposed [Eq.~(\ref{rhos})] has several advantages:
it is trivial to implement and contains a parameter (the density cutoff
$\rho_{\text{c}}$) that gives us control for approaching all-electron
benchmark results. Furthermore, the smoothing procedure is quite general in the
sense that it is basis-set independent. However, the conversion of the potential
from plane waves (obtained as output of the RPS method) to the required format
will depend on the basis set. Nevertheless, most basis sets other
than plane waves use spherical harmonics, such that the use of the
Rayleigh formula for $e^{i\bm{G}\cdot\bm{r}}$ should be the solution in
most codes.

A detailed study of the convergence of the results with respect to the
density and plane-wave expansion cutoffs has shown that quite-well
converged results can be obtained with a plane-wave cutoff that is
reasonable and tremendously smaller than if the all-electron density
was used. Interestingly, a density cutoff $\rho_{\text{c}}=0.3$~Bohr$^{-3}$
seems to be universally good across the whole periodic table of elements.

Then, very well converged results obtained with the
optB88-vdW functional for a large set of strongly bound and vdW solids were
compared to results obtained with other codes based on the PAW or GAPW methods.
Overall, an excellent agreement between the various codes for
the lattice constant and binding energy was obtained.

To conclude, we have shown that it is possible to use the efficient RPS method
within an all-electron framework. Our work should also pave the way for
proposing other methods based on the same idea.
The pseudocharge density method for the Coulomb potential would also be
a possibility, however, one would need to figure out how to calculate the
potential.

\begin{acknowledgments}
This work was supported by the project F41 (SFB ViCoM) of the Austrian Science Fund (FWF).
\end{acknowledgments}

\appendix

\section{\label{appendixA}Derivatives of $\rho_{\text{s}}$}

At $\rho=\rho_{\text{c}}$, a derivative of Eq.~(\ref{rhos}) is continuous
if the corresponding derivatives
for $\rho>\rho_{\text{c}}$ and $\rho\leqslant\rho_{\text{c}}$ are equal.
The first and second derivatives of
$\rho_{\text{s}}$ are given by
\begin{equation}
\nabla\rho_{\text{s}} = \frac{d\rho_{\text{s}}}{d\rho}\nabla\rho
\label{drhos}
\end{equation}
and
\begin{equation}
\nabla^{2}\rho_{\text{s}} =
\frac{d^{2}\rho_{\text{s}}}{d\rho^{2}}\left\vert\nabla\rho\right\vert^{2} +
\frac{d\rho_{\text{s}}}{d\rho}\nabla^{2}\rho,
\label{d2rhos}
\end{equation}
where
\begin{equation}
\frac{d\rho_{\text{s}}}{d\rho} =
\frac{1 + \left(1-n\right)A\left(\rho-\rho_{\text{c}}\right)^{n}}
{\left(1 + A\left(\rho-\rho_{\text{c}}\right)^{n}\right)^{2}}
\label{drhosdrho}
\end{equation}
and
\begin{widetext}
\begin{equation}
\frac{d^{2}\rho_{\text{s}}}{d\rho^{2}} =
-\frac{n(1+n)A\left(\rho-\rho_{\text{c}}\right)^{n-1} +
n(1-n)A^{2}\left(\rho-\rho_{\text{c}}\right)^{2n-1}}
{\left(1 + A\left(\rho-\rho_{\text{c}}\right)^{n}\right)^{3}}.
\label{d2rhosdrho2}
\end{equation}
\end{widetext}
From Eqs.~(\ref{drhos})-(\ref{d2rhosdrho2}), we can see that if $n\geqslant1$,
$\nabla\rho_{\text{s}}$ is continuous since $d\rho_{\text{s}}/d\rho=1$ at
$\rho=\rho_{\text{c}}$. For $\nabla^{2}\rho_{\text{s}}$ it is the case if
$n\geqslant2$ since $d^{2}\rho_{\text{s}}/d\rho^{2}=0$. In general,
for any value of $n\geqslant1$ in Eq.~(\ref{rhos}),
$d\rho_{\text{s}}/d\rho=1$ and $d^{m}\rho_{\text{s}}/d\rho^{m}=0$
for $m=2,3,\ldots,n$ when $\rho=\rho_{\text{c}}$, such that
the $n$ first derivatives of Eq.~(\ref{rhos}) are continuous.

\section{\label{appendixB}Calculation of the potential $v_{\text{c,disp}}^{\text{NL}}$}

In addition of being computationally efficient, the RPS method also leads
to a straightforward calculation of the functional derivative of
Eq.~(\ref{EcdispNL}) [see Eq.~(10) in Ref.~\onlinecite{RomanPerezPRL09}].
However, since in the present case $E_{\text{c,disp}}^{\text{NL}}$ is
evaluated with a density ($\rho_{\text{s}}$) that is not the one ($\rho$) used
to minimize the total energy, the chain rule has
to be applied to get the correct expression for the potential:
\begin{equation}
v_{\text{c,disp}}^{\text{NL}}(\bm{r}) =
\frac{\delta E_{\text{c,disp}}^{\text{NL}}}{\delta\rho(\bm{r})} =
\frac{\delta E_{\text{c,disp}}^{\text{NL}}}{\delta\rho_{\text{s}}(\bm{r})}
\frac{d\rho_{\text{s}}(\bm{r})}{d\rho(\bm{r})},
\label{vcdispNL}
\end{equation}
where $\delta E_{\text{c,disp}}^{\text{NL}}/\delta\rho_{\text{s}}$ is the part
provided by the RPS method and $d\rho_{\text{s}}/d\rho$ is given by Eq.~(\ref{drhosdrho}).
Actually, it should be stressed that it was crucial to use a smoothing scheme
that makes possible the derivation of $d\rho_{\text{s}}/d\rho$.

The second point concerns the conversion of $v_{\text{c,disp}}^{\text{NL}}$
(provided as a pure plane-wave expansion by the RPS procedure) into the LAPW
format, namely (see Sec.~\ref{methodology}), as plane-wave and
spherical harmonics expansions inside the interstitial region and atomic spheres,
respectively:
\begin{equation}
v_{\text{c,disp}}^{\text{NL}}\left(\bm{r}\right) =
\sum_{\bm{G}}v_{\text{c,disp}}^{\text{NL},\bm{G}}
e^{i\bm{G}\cdot\bm{r}},
\label{vcdispNLI}
\end{equation}
\begin{equation}
v_{\text{c,disp}}^{\text{NL}}\left(\bm{r}\right) =
\sum_{\alpha}\sum_{L}\sum_{M=0}^{L}\sum_{p}
v_{\text{c,disp}}^{\text{NL},\alpha LMp}\left(r_{\alpha}\right)
Z_{LMp}\left(\hat{\bm{r}}_{\alpha}\right),
\label{vcdispNLS}
\end{equation}
where $Z_{LMp}$ are real spherical harmonics ($p=\{+,-\}$ for $M\geqslant1$ or
absent for $M=0$) that are defined as follows\cite{KaraAC81}:
\begin{equation}
Z_{L0} =Y_{L0}
\label{ZL0}
\end{equation}
for $M=0$ and
\begin{equation}
Z_{LM+} =
\frac{(-1)^{M}}{\sqrt{2}}\left(Y_{LM} + Y_{LM}^{*}\right),
\label{ZLM+}
\end{equation}
\begin{equation}
Z_{LM-} =
\frac{(-1)^{M}}{i\sqrt{2}}\left(Y_{LM} - Y_{LM}^{*}\right)
\label{ZLM-}
\end{equation}
for $M\geqslant1$.
Of course, the Fourier coefficients $v_{\text{c,disp}}^{\text{NL},\bm{G}}$
in Eq.~(\ref{vcdispNLI})
are the same as those obtained from the RPS method.
The radial functions in Eq.~(\ref{vcdispNLS}) can be obtained
by using the Rayleigh formula ($j_{\ell}$ are spherical Bessel
functions)\cite{Arfken}
\begin{eqnarray}
e^{i\bm{G}\cdot\bm{r}} & = &
4\pi\sum_{\ell=0}^{\infty}\sum_{m=-\ell}^{\ell}
i^{\ell}j_{\ell}\left(Gr\right)
Y_{\ell m}^{*}\left(\widehat{\bm{G}}\right)
Y_{\ell m}\left(\hat{\bm{r}}\right) \nonumber \\
 & = &
4\pi\sum_{L=0}^{\infty}\sum_{M=0}^{L}\sum_{p}
i^{L}j_{L}\left(Gr\right)
Z_{LMp}^{*}\left(\widehat{\bm{G}}\right)
Z_{LMp}\left(\hat{\bm{r}}\right) \nonumber \\
\end{eqnarray}
in Eq.~(\ref{vcdispNLI}), such that
\begin{equation}
v_{\text{c,disp}}^{\text{NL},\alpha LMp}\left(r_{\alpha}\right) =
4\pi i^{L}
\sum_{\bm{G}}v_{\text{c,disp}}^{\text{NL},\bm{G}}
e^{i\bm{G}\cdot\tau_{\alpha}}
j_{L}\left(Gr_{\alpha}\right)
Z_{LMp}^{*}\left(\widehat{\bm{G}}\right),
\end{equation}
where $\tau_{\alpha}$ is the position of nucleus $\alpha$.

\bibliography{/planck/tran/divers/references}

\begin{thebibliography}{77}%
\makeatletter
\providecommand \@ifxundefined [1]{%
 \@ifx{#1\undefined}
}%
\providecommand \@ifnum [1]{%
 \ifnum #1\expandafter \@firstoftwo
 \else \expandafter \@secondoftwo
 \fi
}%
\providecommand \@ifx [1]{%
 \ifx #1\expandafter \@firstoftwo
 \else \expandafter \@secondoftwo
 \fi
}%
\providecommand \natexlab [1]{#1}%
\providecommand \enquote  [1]{``#1''}%
\providecommand \bibnamefont  [1]{#1}%
\providecommand \bibfnamefont [1]{#1}%
\providecommand \citenamefont [1]{#1}%
\providecommand \href@noop [0]{\@secondoftwo}%
\providecommand \href [0]{\begingroup \@sanitize@url \@href}%
\providecommand \@href[1]{\@@startlink{#1}\@@href}%
\providecommand \@@href[1]{\endgroup#1\@@endlink}%
\providecommand \@sanitize@url [0]{\catcode `\\12\catcode `\$12\catcode
  `\&12\catcode `\#12\catcode `\^12\catcode `\_12\catcode `\%12\relax}%
\providecommand \@@startlink[1]{}%
\providecommand \@@endlink[0]{}%
\providecommand \url  [0]{\begingroup\@sanitize@url \@url }%
\providecommand \@url [1]{\endgroup\@href {#1}{\urlprefix }}%
\providecommand \urlprefix  [0]{URL }%
\providecommand \Eprint [0]{\href }%
\providecommand \doibase [0]{http://dx.doi.org/}%
\providecommand \selectlanguage [0]{\@gobble}%
\providecommand \bibinfo  [0]{\@secondoftwo}%
\providecommand \bibfield  [0]{\@secondoftwo}%
\providecommand \translation [1]{[#1]}%
\providecommand \BibitemOpen [0]{}%
\providecommand \bibitemStop [0]{}%
\providecommand \bibitemNoStop [0]{.\EOS\space}%
\providecommand \EOS [0]{\spacefactor3000\relax}%
\providecommand \BibitemShut  [1]{\csname bibitem#1\endcsname}%
\let\auto@bib@innerbib\@empty
\bibitem [{\citenamefont {Hohenberg}\ and\ \citenamefont
  {Kohn}(1964)}]{HohenbergPR64}%
  \BibitemOpen
  \bibfield  {author} {\bibinfo {author} {\bibfnamefont {P.}~\bibnamefont
  {Hohenberg}}\ and\ \bibinfo {author} {\bibfnamefont {W.}~\bibnamefont
  {Kohn}},\ }\href@noop {} {\bibfield  {journal} {\bibinfo  {journal} {Phys.
  Rev.}\ }\textbf {\bibinfo {volume} {136}},\ \bibinfo {pages} {B864} (\bibinfo
  {year} {1964})}\BibitemShut {NoStop}%
\bibitem [{\citenamefont {Kohn}\ and\ \citenamefont {Sham}(1965)}]{KohnPR65}%
  \BibitemOpen
  \bibfield  {author} {\bibinfo {author} {\bibfnamefont {W.}~\bibnamefont
  {Kohn}}\ and\ \bibinfo {author} {\bibfnamefont {L.~J.}\ \bibnamefont
  {Sham}},\ }\href@noop {} {\bibfield  {journal} {\bibinfo  {journal} {Phys.
  Rev.}\ }\textbf {\bibinfo {volume} {140}},\ \bibinfo {pages} {A1133}
  (\bibinfo {year} {1965})}\BibitemShut {NoStop}%
\bibitem [{\citenamefont {Grimme}\ \emph {et~al.}(2016)\citenamefont {Grimme},
  \citenamefont {Hansen}, \citenamefont {Brandenburg},\ and\ \citenamefont
  {Bannwarth}}]{GrimmeCR16}%
  \BibitemOpen
  \bibfield  {author} {\bibinfo {author} {\bibfnamefont {S.}~\bibnamefont
  {Grimme}}, \bibinfo {author} {\bibfnamefont {A.}~\bibnamefont {Hansen}},
  \bibinfo {author} {\bibfnamefont {J.~G.}\ \bibnamefont {Brandenburg}}, \ and\
  \bibinfo {author} {\bibfnamefont {C.}~\bibnamefont {Bannwarth}},\ }\href@noop
  {} {\bibfield  {journal} {\bibinfo  {journal} {Chem. Rev.}\ }\textbf
  {\bibinfo {volume} {116}},\ \bibinfo {pages} {5105} (\bibinfo {year}
  {2016})}\BibitemShut {NoStop}%
\bibitem [{\citenamefont {Becke}(1988)}]{BeckePRA88}%
  \BibitemOpen
  \bibfield  {author} {\bibinfo {author} {\bibfnamefont {A.~D.}\ \bibnamefont
  {Becke}},\ }\href@noop {} {\bibfield  {journal} {\bibinfo  {journal} {Phys.
  Rev. A}\ }\textbf {\bibinfo {volume} {38}},\ \bibinfo {pages} {3098}
  (\bibinfo {year} {1988})}\BibitemShut {NoStop}%
\bibitem [{\citenamefont {Lee}\ \emph {et~al.}(1988)\citenamefont {Lee},
  \citenamefont {Yang},\ and\ \citenamefont {Parr}}]{LeePRB88}%
  \BibitemOpen
  \bibfield  {author} {\bibinfo {author} {\bibfnamefont {C.}~\bibnamefont
  {Lee}}, \bibinfo {author} {\bibfnamefont {W.}~\bibnamefont {Yang}}, \ and\
  \bibinfo {author} {\bibfnamefont {R.~G.}\ \bibnamefont {Parr}},\ }\href@noop
  {} {\bibfield  {journal} {\bibinfo  {journal} {Phys. Rev. B}\ }\textbf
  {\bibinfo {volume} {37}},\ \bibinfo {pages} {785} (\bibinfo {year}
  {1988})}\BibitemShut {NoStop}%
\bibitem [{\citenamefont {Kristy\'{a}n}\ and\ \citenamefont
  {Pulay}(1994)}]{KristyanCPL94}%
  \BibitemOpen
  \bibfield  {author} {\bibinfo {author} {\bibfnamefont {S.}~\bibnamefont
  {Kristy\'{a}n}}\ and\ \bibinfo {author} {\bibfnamefont {P.}~\bibnamefont
  {Pulay}},\ }\href@noop {} {\bibfield  {journal} {\bibinfo  {journal} {Chem.
  Phys. Lett.}\ }\textbf {\bibinfo {volume} {229}},\ \bibinfo {pages} {175}
  (\bibinfo {year} {1994})}\BibitemShut {NoStop}%
\bibitem [{\citenamefont {P\'{e}rez-Jord\'{a}}\ and\ \citenamefont
  {Becke}(1995)}]{PerezJordaCPL95}%
  \BibitemOpen
  \bibfield  {author} {\bibinfo {author} {\bibfnamefont {J.~M.}\ \bibnamefont
  {P\'{e}rez-Jord\'{a}}}\ and\ \bibinfo {author} {\bibfnamefont {A.~D.}\
  \bibnamefont {Becke}},\ }\href@noop {} {\bibfield  {journal} {\bibinfo
  {journal} {Chem. Phys. Lett.}\ }\textbf {\bibinfo {volume} {233}},\ \bibinfo
  {pages} {134} (\bibinfo {year} {1995})}\BibitemShut {NoStop}%
\bibitem [{\citenamefont {Klime{\v{s}}}\ and\ \citenamefont
  {Michaelides}(2012)}]{KlimesJCP12}%
  \BibitemOpen
  \bibfield  {author} {\bibinfo {author} {\bibfnamefont {J.}~\bibnamefont
  {Klime{\v{s}}}}\ and\ \bibinfo {author} {\bibfnamefont {A.}~\bibnamefont
  {Michaelides}},\ }\href@noop {} {\bibfield  {journal} {\bibinfo  {journal}
  {J. Chem. Phys.}\ }\textbf {\bibinfo {volume} {137}},\ \bibinfo {pages}
  {120901} (\bibinfo {year} {2012})}\BibitemShut {NoStop}%
\bibitem [{\citenamefont {Dobson}(2014)}]{DobsonIJQC14}%
  \BibitemOpen
  \bibfield  {author} {\bibinfo {author} {\bibfnamefont {J.~F.}\ \bibnamefont
  {Dobson}},\ }\href@noop {} {\bibfield  {journal} {\bibinfo  {journal} {Int.
  J. Quantum Chem.}\ }\textbf {\bibinfo {volume} {114}},\ \bibinfo {pages}
  {1157} (\bibinfo {year} {2014})}\BibitemShut {NoStop}%
\bibitem [{\citenamefont {Berland}\ \emph {et~al.}(2015)\citenamefont
  {Berland}, \citenamefont {Cooper}, \citenamefont {Lee}, \citenamefont
  {Schr{\"o}der}, \citenamefont {Thonhauser}, \citenamefont {Hyldgaard},\ and\
  \citenamefont {Lundqvist}}]{BerlandRPP15}%
  \BibitemOpen
  \bibfield  {author} {\bibinfo {author} {\bibfnamefont {K.}~\bibnamefont
  {Berland}}, \bibinfo {author} {\bibfnamefont {V.~R.}\ \bibnamefont {Cooper}},
  \bibinfo {author} {\bibfnamefont {K.}~\bibnamefont {Lee}}, \bibinfo {author}
  {\bibfnamefont {E.}~\bibnamefont {Schr{\"o}der}}, \bibinfo {author}
  {\bibfnamefont {T.}~\bibnamefont {Thonhauser}}, \bibinfo {author}
  {\bibfnamefont {P.}~\bibnamefont {Hyldgaard}}, \ and\ \bibinfo {author}
  {\bibfnamefont {B.~I.}\ \bibnamefont {Lundqvist}},\ }\href@noop {} {\bibfield
   {journal} {\bibinfo  {journal} {Rep. Prog. Phys.}\ }\textbf {\bibinfo
  {volume} {78}},\ \bibinfo {pages} {066501} (\bibinfo {year}
  {2015})}\BibitemShut {NoStop}%
\bibitem [{\citenamefont {Wu}\ \emph {et~al.}(2001)\citenamefont {Wu},
  \citenamefont {Vargas}, \citenamefont {Nayak}, \citenamefont {Lotrich},\ and\
  \citenamefont {Scoles}}]{WuJCP01}%
  \BibitemOpen
  \bibfield  {author} {\bibinfo {author} {\bibfnamefont {X.}~\bibnamefont
  {Wu}}, \bibinfo {author} {\bibfnamefont {M.~C.}\ \bibnamefont {Vargas}},
  \bibinfo {author} {\bibfnamefont {S.}~\bibnamefont {Nayak}}, \bibinfo
  {author} {\bibfnamefont {V.}~\bibnamefont {Lotrich}}, \ and\ \bibinfo
  {author} {\bibfnamefont {G.}~\bibnamefont {Scoles}},\ }\href@noop {}
  {\bibfield  {journal} {\bibinfo  {journal} {J. Chem. Phys.}\ }\textbf
  {\bibinfo {volume} {115}},\ \bibinfo {pages} {8748} (\bibinfo {year}
  {2001})}\BibitemShut {NoStop}%
\bibitem [{\citenamefont {Wu}\ and\ \citenamefont {Yang}(2002)}]{WuJCP02}%
  \BibitemOpen
  \bibfield  {author} {\bibinfo {author} {\bibfnamefont {Q.}~\bibnamefont
  {Wu}}\ and\ \bibinfo {author} {\bibfnamefont {W.}~\bibnamefont {Yang}},\
  }\href@noop {} {\bibfield  {journal} {\bibinfo  {journal} {J. Chem. Phys.}\
  }\textbf {\bibinfo {volume} {116}},\ \bibinfo {pages} {515} (\bibinfo {year}
  {2002})}\BibitemShut {NoStop}%
\bibitem [{\citenamefont {Hasegawa}\ and\ \citenamefont
  {Nishidate}(2004)}]{HasegawaPRB04}%
  \BibitemOpen
  \bibfield  {author} {\bibinfo {author} {\bibfnamefont {M.}~\bibnamefont
  {Hasegawa}}\ and\ \bibinfo {author} {\bibfnamefont {K.}~\bibnamefont
  {Nishidate}},\ }\href@noop {} {\bibfield  {journal} {\bibinfo  {journal}
  {Phys. Rev. B}\ }\textbf {\bibinfo {volume} {70}},\ \bibinfo {pages} {205431}
  (\bibinfo {year} {2004})}\BibitemShut {NoStop}%
\bibitem [{\citenamefont {Grimme}(2004)}]{GrimmeJCC04}%
  \BibitemOpen
  \bibfield  {author} {\bibinfo {author} {\bibfnamefont {S.}~\bibnamefont
  {Grimme}},\ }\href@noop {} {\bibfield  {journal} {\bibinfo  {journal} {J.
  Comput. Chem.}\ }\textbf {\bibinfo {volume} {25}},\ \bibinfo {pages} {1463}
  (\bibinfo {year} {2004})}\BibitemShut {NoStop}%
\bibitem [{\citenamefont {Becke}\ and\ \citenamefont
  {Johnson}(2005)}]{BeckeJCP05}%
  \BibitemOpen
  \bibfield  {author} {\bibinfo {author} {\bibfnamefont {A.~D.}\ \bibnamefont
  {Becke}}\ and\ \bibinfo {author} {\bibfnamefont {E.~R.}\ \bibnamefont
  {Johnson}},\ }\href@noop {} {\bibfield  {journal} {\bibinfo  {journal} {J.
  Chem. Phys.}\ }\textbf {\bibinfo {volume} {122}},\ \bibinfo {pages} {154104}
  (\bibinfo {year} {2005})}\BibitemShut {NoStop}%
\bibitem [{\citenamefont {Tkatchenko}\ and\ \citenamefont
  {Scheffler}(2009)}]{TkatchenkoPRL09}%
  \BibitemOpen
  \bibfield  {author} {\bibinfo {author} {\bibfnamefont {A.}~\bibnamefont
  {Tkatchenko}}\ and\ \bibinfo {author} {\bibfnamefont {M.}~\bibnamefont
  {Scheffler}},\ }\href@noop {} {\bibfield  {journal} {\bibinfo  {journal}
  {Phys. Rev. Lett.}\ }\textbf {\bibinfo {volume} {102}},\ \bibinfo {pages}
  {073005} (\bibinfo {year} {2009})}\BibitemShut {NoStop}%
\bibitem [{\citenamefont {Grimme}\ \emph {et~al.}(2010)\citenamefont {Grimme},
  \citenamefont {Antony}, \citenamefont {Ehrlich},\ and\ \citenamefont
  {Krieg}}]{GrimmeJCP10}%
  \BibitemOpen
  \bibfield  {author} {\bibinfo {author} {\bibfnamefont {S.}~\bibnamefont
  {Grimme}}, \bibinfo {author} {\bibfnamefont {J.}~\bibnamefont {Antony}},
  \bibinfo {author} {\bibfnamefont {S.}~\bibnamefont {Ehrlich}}, \ and\
  \bibinfo {author} {\bibfnamefont {H.}~\bibnamefont {Krieg}},\ }\href@noop {}
  {\bibfield  {journal} {\bibinfo  {journal} {J. Chem. Phys.}\ }\textbf
  {\bibinfo {volume} {132}},\ \bibinfo {pages} {154104} (\bibinfo {year}
  {2010})}\BibitemShut {NoStop}%
\bibitem [{\citenamefont {Grimme}\ \emph {et~al.}(2011)\citenamefont {Grimme},
  \citenamefont {Ehrlich},\ and\ \citenamefont {Goerigk}}]{GrimmeJCC11}%
  \BibitemOpen
  \bibfield  {author} {\bibinfo {author} {\bibfnamefont {S.}~\bibnamefont
  {Grimme}}, \bibinfo {author} {\bibfnamefont {S.}~\bibnamefont {Ehrlich}}, \
  and\ \bibinfo {author} {\bibfnamefont {L.}~\bibnamefont {Goerigk}},\
  }\href@noop {} {\bibfield  {journal} {\bibinfo  {journal} {J. Comput. Chem.}\
  }\textbf {\bibinfo {volume} {32}},\ \bibinfo {pages} {1456} (\bibinfo {year}
  {2011})}\BibitemShut {NoStop}%
\bibitem [{\citenamefont {Tkatchenko}\ \emph {et~al.}(2012)\citenamefont
  {Tkatchenko}, \citenamefont {DiStasio}, \citenamefont {Car},\ and\
  \citenamefont {Scheffler}}]{TkatchenkoPRL12}%
  \BibitemOpen
  \bibfield  {author} {\bibinfo {author} {\bibfnamefont {A.}~\bibnamefont
  {Tkatchenko}}, \bibinfo {author} {\bibfnamefont {R.~A.}\ \bibnamefont
  {DiStasio}, \bibfnamefont {Jr.}}, \bibinfo {author} {\bibfnamefont
  {R.}~\bibnamefont {Car}}, \ and\ \bibinfo {author} {\bibfnamefont
  {M.}~\bibnamefont {Scheffler}},\ }\href@noop {} {\bibfield  {journal}
  {\bibinfo  {journal} {Phys. Rev. Lett.}\ }\textbf {\bibinfo {volume} {108}},\
  \bibinfo {pages} {236402} (\bibinfo {year} {2012})}\BibitemShut {NoStop}%
\bibitem [{\citenamefont {Dion}\ \emph {et~al.}(2004)\citenamefont {Dion},
  \citenamefont {Rydberg}, \citenamefont {Schr\"oder}, \citenamefont
  {Langreth},\ and\ \citenamefont {Lundqvist}}]{DionPRL04}%
  \BibitemOpen
  \bibfield  {author} {\bibinfo {author} {\bibfnamefont {M.}~\bibnamefont
  {Dion}}, \bibinfo {author} {\bibfnamefont {H.}~\bibnamefont {Rydberg}},
  \bibinfo {author} {\bibfnamefont {E.}~\bibnamefont {Schr\"oder}}, \bibinfo
  {author} {\bibfnamefont {D.~C.}\ \bibnamefont {Langreth}}, \ and\ \bibinfo
  {author} {\bibfnamefont {B.~I.}\ \bibnamefont {Lundqvist}},\ }\href@noop {}
  {\bibfield  {journal} {\bibinfo  {journal} {Phys. Rev. Lett.}\ }\textbf
  {\bibinfo {volume} {92}},\ \bibinfo {pages} {246401} (\bibinfo {year}
  {2004})};\ \bibinfo {note} {\textbf{95}, 109902(E) (2005)}\BibitemShut
  {NoStop}%
\bibitem [{\citenamefont {Langreth}\ \emph {et~al.}(2009)\citenamefont
  {Langreth}, \citenamefont {Lundqvist}, \citenamefont {Chakarova-K\"{a}ck},
  \citenamefont {Cooper}, \citenamefont {Dion}, \citenamefont {Hyldgaard},
  \citenamefont {Kelkkanen}, \citenamefont {Kleis}, \citenamefont {Kong},
  \citenamefont {Li}, \citenamefont {Moses}, \citenamefont {Murray},
  \citenamefont {Puzder}, \citenamefont {Rydberg}, \citenamefont
  {Schr\"{o}der},\ and\ \citenamefont {Thonhauser}}]{LangrethJPCM09}%
  \BibitemOpen
  \bibfield  {author} {\bibinfo {author} {\bibfnamefont {D.~C.}\ \bibnamefont
  {Langreth}}, \bibinfo {author} {\bibfnamefont {B.~I.}\ \bibnamefont
  {Lundqvist}}, \bibinfo {author} {\bibfnamefont {S.~D.}\ \bibnamefont
  {Chakarova-K\"{a}ck}}, \bibinfo {author} {\bibfnamefont {V.~R.}\ \bibnamefont
  {Cooper}}, \bibinfo {author} {\bibfnamefont {M.}~\bibnamefont {Dion}},
  \bibinfo {author} {\bibfnamefont {P.}~\bibnamefont {Hyldgaard}}, \bibinfo
  {author} {\bibfnamefont {A.}~\bibnamefont {Kelkkanen}}, \bibinfo {author}
  {\bibfnamefont {J.}~\bibnamefont {Kleis}}, \bibinfo {author} {\bibfnamefont
  {L.}~\bibnamefont {Kong}}, \bibinfo {author} {\bibfnamefont {S.}~\bibnamefont
  {Li}}, \bibinfo {author} {\bibfnamefont {P.~G.}\ \bibnamefont {Moses}},
  \bibinfo {author} {\bibfnamefont {E.}~\bibnamefont {Murray}}, \bibinfo
  {author} {\bibfnamefont {A.}~\bibnamefont {Puzder}}, \bibinfo {author}
  {\bibfnamefont {H.}~\bibnamefont {Rydberg}}, \bibinfo {author} {\bibfnamefont
  {E.}~\bibnamefont {Schr\"{o}der}}, \ and\ \bibinfo {author} {\bibfnamefont
  {T.}~\bibnamefont {Thonhauser}},\ }\href@noop {} {\bibfield  {journal}
  {\bibinfo  {journal} {J. Phys.: Condens. Matter}\ }\textbf {\bibinfo {volume}
  {21}},\ \bibinfo {pages} {084203} (\bibinfo {year} {2009})}\BibitemShut
  {NoStop}%
\bibitem [{\citenamefont {Gunnarsson}\ and\ \citenamefont
  {Lundqvist}(1976)}]{GunnarssonPRB76}%
  \BibitemOpen
  \bibfield  {author} {\bibinfo {author} {\bibfnamefont {O.}~\bibnamefont
  {Gunnarsson}}\ and\ \bibinfo {author} {\bibfnamefont {B.~I.}\ \bibnamefont
  {Lundqvist}},\ }\href@noop {} {\bibfield  {journal} {\bibinfo  {journal}
  {Phys. Rev. B}\ }\textbf {\bibinfo {volume} {13}},\ \bibinfo {pages} {4274}
  (\bibinfo {year} {1976})},\ \bibinfo {note} {\textbf{15}, 6006
  (1977)}\BibitemShut {NoStop}%
\bibitem [{\citenamefont {Langreth}\ and\ \citenamefont
  {Perdew}(1977)}]{LangrethPRB77}%
  \BibitemOpen
  \bibfield  {author} {\bibinfo {author} {\bibfnamefont {D.~C.}\ \bibnamefont
  {Langreth}}\ and\ \bibinfo {author} {\bibfnamefont {J.~P.}\ \bibnamefont
  {Perdew}},\ }\href@noop {} {\bibfield  {journal} {\bibinfo  {journal} {Phys.
  Rev. B}\ }\textbf {\bibinfo {volume} {15}},\ \bibinfo {pages} {2884}
  (\bibinfo {year} {1977})}\BibitemShut {NoStop}%
\bibitem [{\citenamefont {Zhang}\ and\ \citenamefont
  {Yang}(1998)}]{ZhangPRL98}%
  \BibitemOpen
  \bibfield  {author} {\bibinfo {author} {\bibfnamefont {Y.}~\bibnamefont
  {Zhang}}\ and\ \bibinfo {author} {\bibfnamefont {W.}~\bibnamefont {Yang}},\
  }\href@noop {} {\bibfield  {journal} {\bibinfo  {journal} {Phys. Rev. Lett.}\
  }\textbf {\bibinfo {volume} {80}},\ \bibinfo {pages} {890} (\bibinfo {year}
  {1998})}\BibitemShut {NoStop}%
\bibitem [{\citenamefont {Vosko}\ \emph {et~al.}(1980)\citenamefont {Vosko},
  \citenamefont {Wilk},\ and\ \citenamefont {Nusair}}]{VoskoCJP80}%
  \BibitemOpen
  \bibfield  {author} {\bibinfo {author} {\bibfnamefont {S.~H.}\ \bibnamefont
  {Vosko}}, \bibinfo {author} {\bibfnamefont {L.}~\bibnamefont {Wilk}}, \ and\
  \bibinfo {author} {\bibfnamefont {M.}~\bibnamefont {Nusair}},\ }\href@noop {}
  {\bibfield  {journal} {\bibinfo  {journal} {Can. J. Phys.}\ }\textbf
  {\bibinfo {volume} {58}},\ \bibinfo {pages} {1200} (\bibinfo {year}
  {1980})}\BibitemShut {NoStop}%
\bibitem [{\citenamefont {Perdew}\ and\ \citenamefont
  {Wang}(1992)}]{PerdewPRB92a}%
  \BibitemOpen
  \bibfield  {author} {\bibinfo {author} {\bibfnamefont {J.~P.}\ \bibnamefont
  {Perdew}}\ and\ \bibinfo {author} {\bibfnamefont {Y.}~\bibnamefont {Wang}},\
  }\href@noop {} {\bibfield  {journal} {\bibinfo  {journal} {Phys. Rev. B}\
  }\textbf {\bibinfo {volume} {45}},\ \bibinfo {pages} {13244} (\bibinfo {year}
  {1992})}\BibitemShut {NoStop}%
\bibitem [{\citenamefont {Lee}\ \emph {et~al.}(2010)\citenamefont {Lee},
  \citenamefont {Murray}, \citenamefont {Kong}, \citenamefont {Lundqvist},\
  and\ \citenamefont {Langreth}}]{LeePRB10}%
  \BibitemOpen
  \bibfield  {author} {\bibinfo {author} {\bibfnamefont {K.}~\bibnamefont
  {Lee}}, \bibinfo {author} {\bibfnamefont {E.~D.}\ \bibnamefont {Murray}},
  \bibinfo {author} {\bibfnamefont {L.}~\bibnamefont {Kong}}, \bibinfo {author}
  {\bibfnamefont {B.~I.}\ \bibnamefont {Lundqvist}}, \ and\ \bibinfo {author}
  {\bibfnamefont {D.~C.}\ \bibnamefont {Langreth}},\ }\href@noop {} {\bibfield
  {journal} {\bibinfo  {journal} {Phys. Rev. B}\ }\textbf {\bibinfo {volume}
  {82}},\ \bibinfo {pages} {081101(R)} (\bibinfo {year} {2010})}\BibitemShut
  {NoStop}%
\bibitem [{\citenamefont {Klime{\v{s}}}\ \emph {et~al.}(2010)\citenamefont
  {Klime{\v{s}}}, \citenamefont {Bowler},\ and\ \citenamefont
  {Michaelides}}]{KlimesJPCM10}%
  \BibitemOpen
  \bibfield  {author} {\bibinfo {author} {\bibfnamefont {J.}~\bibnamefont
  {Klime{\v{s}}}}, \bibinfo {author} {\bibfnamefont {D.~R.}\ \bibnamefont
  {Bowler}}, \ and\ \bibinfo {author} {\bibfnamefont {A.}~\bibnamefont
  {Michaelides}},\ }\href@noop {} {\bibfield  {journal} {\bibinfo  {journal}
  {J. Phys.: Condens. Matter}\ }\textbf {\bibinfo {volume} {22}},\ \bibinfo
  {pages} {022201} (\bibinfo {year} {2010})}\BibitemShut {NoStop}%
\bibitem [{\citenamefont {Cooper}(2010)}]{CooperPRB10}%
  \BibitemOpen
  \bibfield  {author} {\bibinfo {author} {\bibfnamefont {V.~R.}\ \bibnamefont
  {Cooper}},\ }\href@noop {} {\bibfield  {journal} {\bibinfo  {journal} {Phys.
  Rev. B}\ }\textbf {\bibinfo {volume} {81}},\ \bibinfo {pages} {161104(R)}
  (\bibinfo {year} {2010})}\BibitemShut {NoStop}%
\bibitem [{\citenamefont {Vydrov}\ and\ \citenamefont
  {Van~Voorhis}(2009)}]{VydrovPRL09}%
  \BibitemOpen
  \bibfield  {author} {\bibinfo {author} {\bibfnamefont {O.~A.}\ \bibnamefont
  {Vydrov}}\ and\ \bibinfo {author} {\bibfnamefont {T.}~\bibnamefont
  {Van~Voorhis}},\ }\href@noop {} {\bibfield  {journal} {\bibinfo  {journal}
  {Phys. Rev. Lett.}\ }\textbf {\bibinfo {volume} {103}},\ \bibinfo {pages}
  {063004} (\bibinfo {year} {2009})}\BibitemShut {NoStop}%
\bibitem [{\citenamefont {Vydrov}\ and\ \citenamefont
  {Van~Voorhis}(2010)}]{VydrovJCP10}%
  \BibitemOpen
  \bibfield  {author} {\bibinfo {author} {\bibfnamefont {O.~A.}\ \bibnamefont
  {Vydrov}}\ and\ \bibinfo {author} {\bibfnamefont {T.}~\bibnamefont
  {Van~Voorhis}},\ }\href@noop {} {\bibfield  {journal} {\bibinfo  {journal}
  {J. Chem. Phys.}\ }\textbf {\bibinfo {volume} {133}},\ \bibinfo {pages}
  {244103} (\bibinfo {year} {2010})}\BibitemShut {NoStop}%
\bibitem [{\citenamefont {Wellendorff}\ \emph {et~al.}(2010)\citenamefont
  {Wellendorff}, \citenamefont {Kelkkanen}, \citenamefont {Mortensen},
  \citenamefont {Lundqvist},\ and\ \citenamefont {Bligaard}}]{WellendorffTC10}%
  \BibitemOpen
  \bibfield  {author} {\bibinfo {author} {\bibfnamefont {J.}~\bibnamefont
  {Wellendorff}}, \bibinfo {author} {\bibfnamefont {A.}~\bibnamefont
  {Kelkkanen}}, \bibinfo {author} {\bibfnamefont {J.~J.}\ \bibnamefont
  {Mortensen}}, \bibinfo {author} {\bibfnamefont {B.~I.}\ \bibnamefont
  {Lundqvist}}, \ and\ \bibinfo {author} {\bibfnamefont {T.}~\bibnamefont
  {Bligaard}},\ }\href@noop {} {\bibfield  {journal} {\bibinfo  {journal} {Top.
  Catal.}\ }\textbf {\bibinfo {volume} {53}},\ \bibinfo {pages} {378} (\bibinfo
  {year} {2010})}\BibitemShut {NoStop}%
\bibitem [{\citenamefont {Sabatini}\ \emph {et~al.}(2013)\citenamefont
  {Sabatini}, \citenamefont {Gorni},\ and\ \citenamefont
  {de~Gironcoli}}]{SabatiniPRB13}%
  \BibitemOpen
  \bibfield  {author} {\bibinfo {author} {\bibfnamefont {R.}~\bibnamefont
  {Sabatini}}, \bibinfo {author} {\bibfnamefont {T.}~\bibnamefont {Gorni}}, \
  and\ \bibinfo {author} {\bibfnamefont {S.}~\bibnamefont {de~Gironcoli}},\
  }\href@noop {} {\bibfield  {journal} {\bibinfo  {journal} {Phys. Rev. B}\
  }\textbf {\bibinfo {volume} {87}},\ \bibinfo {pages} {041108(R)} (\bibinfo
  {year} {2013})}\BibitemShut {NoStop}%
\bibitem [{\citenamefont {Hamada}(2014)}]{HamadaPRB14}%
  \BibitemOpen
  \bibfield  {author} {\bibinfo {author} {\bibfnamefont {I.}~\bibnamefont
  {Hamada}},\ }\href@noop {} {\bibfield  {journal} {\bibinfo  {journal} {Phys.
  Rev. B}\ }\textbf {\bibinfo {volume} {89}},\ \bibinfo {pages} {121103(R)}
  (\bibinfo {year} {2014})}\BibitemShut {NoStop}%
\bibitem [{\citenamefont {Berland}\ and\ \citenamefont
  {Hyldgaard}(2014)}]{BerlandPRB14}%
  \BibitemOpen
  \bibfield  {author} {\bibinfo {author} {\bibfnamefont {K.}~\bibnamefont
  {Berland}}\ and\ \bibinfo {author} {\bibfnamefont {P.}~\bibnamefont
  {Hyldgaard}},\ }\href@noop {} {\bibfield  {journal} {\bibinfo  {journal}
  {Phys. Rev. B}\ }\textbf {\bibinfo {volume} {89}},\ \bibinfo {pages} {035412}
  (\bibinfo {year} {2014})}\BibitemShut {NoStop}%
\bibitem [{\citenamefont {Peng}\ \emph {et~al.}(2016)\citenamefont {Peng},
  \citenamefont {Yang}, \citenamefont {Perdew},\ and\ \citenamefont
  {Sun}}]{PengPRX16}%
  \BibitemOpen
  \bibfield  {author} {\bibinfo {author} {\bibfnamefont {H.}~\bibnamefont
  {Peng}}, \bibinfo {author} {\bibfnamefont {Z.-H.}\ \bibnamefont {Yang}},
  \bibinfo {author} {\bibfnamefont {J.~P.}\ \bibnamefont {Perdew}}, \ and\
  \bibinfo {author} {\bibfnamefont {J.}~\bibnamefont {Sun}},\ }\href@noop {}
  {\bibfield  {journal} {\bibinfo  {journal} {Phys. Rev. X}\ }\textbf {\bibinfo
  {volume} {6}},\ \bibinfo {pages} {041005} (\bibinfo {year}
  {2016})}\BibitemShut {NoStop}%
\bibitem [{\citenamefont {Grimme}(2011)}]{GrimmeWCMS11}%
  \BibitemOpen
  \bibfield  {author} {\bibinfo {author} {\bibfnamefont {S.}~\bibnamefont
  {Grimme}},\ }\href@noop {} {\bibfield  {journal} {\bibinfo  {journal} {WIREs
  Comput. Mol. Sci.}\ }\textbf {\bibinfo {volume} {1}},\ \bibinfo {pages} {211}
  (\bibinfo {year} {2011})}\BibitemShut {NoStop}%
\bibitem [{\citenamefont {Goerigk}(2014)}]{GoerigkJCT10}%
  \BibitemOpen
  \bibfield  {author} {\bibinfo {author} {\bibfnamefont {L.}~\bibnamefont
  {Goerigk}},\ }\href@noop {} {\bibfield  {journal} {\bibinfo  {journal} {J.
  Chem. Theory Comput.}\ }\textbf {\bibinfo {volume} {10}},\ \bibinfo {pages}
  {968} (\bibinfo {year} {2014})}\BibitemShut {NoStop}%
\bibitem [{\citenamefont {R\^{e}go}\ \emph {et~al.}(2015)\citenamefont
  {R\^{e}go}, \citenamefont {Oliveira}, \citenamefont {Tereshchuk},\ and\
  \citenamefont {Da~Silva}}]{RegoJPCM15}%
  \BibitemOpen
  \bibfield  {author} {\bibinfo {author} {\bibfnamefont {C.~R.~C.}\
  \bibnamefont {R\^{e}go}}, \bibinfo {author} {\bibfnamefont {L.~N.}\
  \bibnamefont {Oliveira}}, \bibinfo {author} {\bibfnamefont {P.}~\bibnamefont
  {Tereshchuk}}, \ and\ \bibinfo {author} {\bibfnamefont {J.~L.~F.}\
  \bibnamefont {Da~Silva}},\ }\href@noop {} {\bibfield  {journal} {\bibinfo
  {journal} {J. Phys.: Condens. Matter}\ }\textbf {\bibinfo {volume} {27}},\
  \bibinfo {pages} {415502} (\bibinfo {year} {2015})};\ \bibinfo {note}
  {\textbf{28}, 129501 (2016)}\BibitemShut {NoStop}%
\bibitem [{\citenamefont {Tran}\ \emph {et~al.}(2016)\citenamefont {Tran},
  \citenamefont {Stelzl},\ and\ \citenamefont {Blaha}}]{TranJCP16}%
  \BibitemOpen
  \bibfield  {author} {\bibinfo {author} {\bibfnamefont {F.}~\bibnamefont
  {Tran}}, \bibinfo {author} {\bibfnamefont {J.}~\bibnamefont {Stelzl}}, \ and\
  \bibinfo {author} {\bibfnamefont {P.}~\bibnamefont {Blaha}},\ }\href@noop {}
  {\bibfield  {journal} {\bibinfo  {journal} {J. Chem. Phys.}\ }\textbf
  {\bibinfo {volume} {144}},\ \bibinfo {pages} {204120} (\bibinfo {year}
  {2016})}\BibitemShut {NoStop}%
\bibitem [{\citenamefont {Lozano}\ \emph {et~al.}(2017)\citenamefont {Lozano},
  \citenamefont {Escribano}, \citenamefont {Akhmatskaya},\ and\ \citenamefont
  {Carrasco}}]{LozanoPCCP17}%
  \BibitemOpen
  \bibfield  {author} {\bibinfo {author} {\bibfnamefont {A.}~\bibnamefont
  {Lozano}}, \bibinfo {author} {\bibfnamefont {B.}~\bibnamefont {Escribano}},
  \bibinfo {author} {\bibfnamefont {E.}~\bibnamefont {Akhmatskaya}}, \ and\
  \bibinfo {author} {\bibfnamefont {J.}~\bibnamefont {Carrasco}},\ }\href@noop
  {} {\bibfield  {journal} {\bibinfo  {journal} {Phys. Chem. Chem. Phys.}\
  }\textbf {\bibinfo {volume} {19}},\ \bibinfo {pages} {10133} (\bibinfo {year}
  {2017})}\BibitemShut {NoStop}%
\bibitem [{\citenamefont {Lazi\'{c}}\ \emph {et~al.}(2010)\citenamefont
  {Lazi\'{c}}, \citenamefont {Atodiresei}, \citenamefont {Alaei}, \citenamefont
  {Caciuc}, \citenamefont {Bl\"{u}gel},\ and\ \citenamefont
  {Brako}}]{LazicCPC10}%
  \BibitemOpen
  \bibfield  {author} {\bibinfo {author} {\bibfnamefont {P.}~\bibnamefont
  {Lazi\'{c}}}, \bibinfo {author} {\bibfnamefont {N.}~\bibnamefont
  {Atodiresei}}, \bibinfo {author} {\bibfnamefont {M.}~\bibnamefont {Alaei}},
  \bibinfo {author} {\bibfnamefont {V.}~\bibnamefont {Caciuc}}, \bibinfo
  {author} {\bibfnamefont {S.}~\bibnamefont {Bl\"{u}gel}}, \ and\ \bibinfo
  {author} {\bibfnamefont {R.}~\bibnamefont {Brako}},\ }\href@noop {}
  {\bibfield  {journal} {\bibinfo  {journal} {Comput. Phys. Commun.}\ }\textbf
  {\bibinfo {volume} {181}},\ \bibinfo {pages} {371} (\bibinfo {year}
  {2010})}\BibitemShut {NoStop}%
\bibitem [{\citenamefont {Nabok}\ \emph {et~al.}(2011)\citenamefont {Nabok},
  \citenamefont {Puschnig},\ and\ \citenamefont {Ambrosch-Draxl}}]{NabokCPC11}%
  \BibitemOpen
  \bibfield  {author} {\bibinfo {author} {\bibfnamefont {D.}~\bibnamefont
  {Nabok}}, \bibinfo {author} {\bibfnamefont {P.}~\bibnamefont {Puschnig}}, \
  and\ \bibinfo {author} {\bibfnamefont {C.}~\bibnamefont {Ambrosch-Draxl}},\
  }\href@noop {} {\bibfield  {journal} {\bibinfo  {journal} {Comput. Phys.
  Commun.}\ }\textbf {\bibinfo {volume} {182}},\ \bibinfo {pages} {1657}
  (\bibinfo {year} {2011})}\BibitemShut {NoStop}%
\bibitem [{\citenamefont {Rom\'an-P\'erez}\ and\ \citenamefont
  {Soler}(2009)}]{RomanPerezPRL09}%
  \BibitemOpen
  \bibfield  {author} {\bibinfo {author} {\bibfnamefont {G.}~\bibnamefont
  {Rom\'an-P\'erez}}\ and\ \bibinfo {author} {\bibfnamefont {J.~M.}\
  \bibnamefont {Soler}},\ }\href@noop {} {\bibfield  {journal} {\bibinfo
  {journal} {Phys. Rev. Lett.}\ }\textbf {\bibinfo {volume} {103}},\ \bibinfo
  {pages} {096102} (\bibinfo {year} {2009})}\BibitemShut {NoStop}%
\bibitem [{\citenamefont {Sabatini}\ \emph {et~al.}(2012)\citenamefont
  {Sabatini}, \citenamefont {K\"{u}\c{c}\"{u}kbenli}, \citenamefont {Kolb},
  \citenamefont {Thonhauser},\ and\ \citenamefont
  {de~Gironcoli}}]{SabatiniJPCM12}%
  \BibitemOpen
  \bibfield  {author} {\bibinfo {author} {\bibfnamefont {R.}~\bibnamefont
  {Sabatini}}, \bibinfo {author} {\bibfnamefont {E.}~\bibnamefont
  {K\"{u}\c{c}\"{u}kbenli}}, \bibinfo {author} {\bibfnamefont {B.}~\bibnamefont
  {Kolb}}, \bibinfo {author} {\bibfnamefont {T.}~\bibnamefont {Thonhauser}}, \
  and\ \bibinfo {author} {\bibfnamefont {S.}~\bibnamefont {de~Gironcoli}},\
  }\href@noop {} {\bibfield  {journal} {\bibinfo  {journal} {J. Phys.: Condens.
  Matter}\ }\textbf {\bibinfo {volume} {24}},\ \bibinfo {pages} {424209}
  (\bibinfo {year} {2012})}\BibitemShut {NoStop}%
\bibitem [{\citenamefont {Klime{\v{s}}}\ \emph {et~al.}(2011)\citenamefont
  {Klime{\v{s}}}, \citenamefont {Bowler},\ and\ \citenamefont
  {Michaelides}}]{KlimesPRB11}%
  \BibitemOpen
  \bibfield  {author} {\bibinfo {author} {\bibfnamefont {J.}~\bibnamefont
  {Klime{\v{s}}}}, \bibinfo {author} {\bibfnamefont {D.~R.}\ \bibnamefont
  {Bowler}}, \ and\ \bibinfo {author} {\bibfnamefont {A.}~\bibnamefont
  {Michaelides}},\ }\href@noop {} {\bibfield  {journal} {\bibinfo  {journal}
  {Phys. Rev. B}\ }\textbf {\bibinfo {volume} {83}},\ \bibinfo {pages} {195131}
  (\bibinfo {year} {2011})}\BibitemShut {NoStop}%
\bibitem [{\citenamefont {Tran}\ and\ \citenamefont
  {Hutter}(2013)}]{TranJCP13}%
  \BibitemOpen
  \bibfield  {author} {\bibinfo {author} {\bibfnamefont {F.}~\bibnamefont
  {Tran}}\ and\ \bibinfo {author} {\bibfnamefont {J.}~\bibnamefont {Hutter}},\
  }\href@noop {} {\bibfield  {journal} {\bibinfo  {journal} {J. Chem. Phys.}\
  }\textbf {\bibinfo {volume} {138}},\ \bibinfo {pages} {204103} (\bibinfo
  {year} {2013})};\ \bibinfo {note} {\textbf{139}, 039903 (2013)}\BibitemShut
  {NoStop}%
\bibitem [{\citenamefont {Larsen}\ \emph {et~al.}(2017)\citenamefont {Larsen},
  \citenamefont {Kuisma}, \citenamefont {L\"{o}fgren}, \citenamefont
  {Pouillon}, \citenamefont {Erhart},\ and\ \citenamefont
  {Hyldgaard}}]{LarsenMSMSE17}%
  \BibitemOpen
  \bibfield  {author} {\bibinfo {author} {\bibfnamefont {A.~H.}\ \bibnamefont
  {Larsen}}, \bibinfo {author} {\bibfnamefont {M.}~\bibnamefont {Kuisma}},
  \bibinfo {author} {\bibfnamefont {J.}~\bibnamefont {L\"{o}fgren}}, \bibinfo
  {author} {\bibfnamefont {Y.}~\bibnamefont {Pouillon}}, \bibinfo {author}
  {\bibfnamefont {P.}~\bibnamefont {Erhart}}, \ and\ \bibinfo {author}
  {\bibfnamefont {P.}~\bibnamefont {Hyldgaard}},\ }\href@noop {} {\bibfield
  {journal} {\bibinfo  {journal} {Modelling Simul. Mater. Sci. Eng.}\ }\textbf
  {\bibinfo {volume} {25}},\ \bibinfo {pages} {065004} (\bibinfo {year}
  {2017})}\BibitemShut {NoStop}%
\bibitem [{\citenamefont {Thonhauser}\ \emph {et~al.}(2007)\citenamefont
  {Thonhauser}, \citenamefont {Cooper}, \citenamefont {Li}, \citenamefont
  {Puzder}, \citenamefont {Hyldgaard},\ and\ \citenamefont
  {Langreth}}]{ThonhauserPRB07}%
  \BibitemOpen
  \bibfield  {author} {\bibinfo {author} {\bibfnamefont {T.}~\bibnamefont
  {Thonhauser}}, \bibinfo {author} {\bibfnamefont {V.~R.}\ \bibnamefont
  {Cooper}}, \bibinfo {author} {\bibfnamefont {S.}~\bibnamefont {Li}}, \bibinfo
  {author} {\bibfnamefont {A.}~\bibnamefont {Puzder}}, \bibinfo {author}
  {\bibfnamefont {P.}~\bibnamefont {Hyldgaard}}, \ and\ \bibinfo {author}
  {\bibfnamefont {D.~C.}\ \bibnamefont {Langreth}},\ }\href@noop {} {\bibfield
  {journal} {\bibinfo  {journal} {Phys. Rev. B}\ }\textbf {\bibinfo {volume}
  {76}},\ \bibinfo {pages} {125112} (\bibinfo {year} {2007})}\BibitemShut
  {NoStop}%
\bibitem [{\citenamefont {Gulans}\ \emph {et~al.}(2009)\citenamefont {Gulans},
  \citenamefont {Puska},\ and\ \citenamefont {Nieminen}}]{GulansPRB09}%
  \BibitemOpen
  \bibfield  {author} {\bibinfo {author} {\bibfnamefont {A.}~\bibnamefont
  {Gulans}}, \bibinfo {author} {\bibfnamefont {M.~J.}\ \bibnamefont {Puska}}, \
  and\ \bibinfo {author} {\bibfnamefont {R.~M.}\ \bibnamefont {Nieminen}},\
  }\href@noop {} {\bibfield  {journal} {\bibinfo  {journal} {Phys. Rev. B}\
  }\textbf {\bibinfo {volume} {79}},\ \bibinfo {pages} {201105(R)} (\bibinfo
  {year} {2009})}\BibitemShut {NoStop}%
\bibitem [{\citenamefont {Andersen}(1975)}]{AndersenPRB75}%
  \BibitemOpen
  \bibfield  {author} {\bibinfo {author} {\bibfnamefont {O.~K.}\ \bibnamefont
  {Andersen}},\ }\href@noop {} {\bibfield  {journal} {\bibinfo  {journal}
  {Phys. Rev. B}\ }\textbf {\bibinfo {volume} {12}},\ \bibinfo {pages} {3060}
  (\bibinfo {year} {1975})}\BibitemShut {NoStop}%
\bibitem [{\citenamefont {Singh}\ and\ \citenamefont
  {Nordstr{\"{o}}m}(2006)}]{Singh}%
  \BibitemOpen
  \bibfield  {author} {\bibinfo {author} {\bibfnamefont {D.~J.}\ \bibnamefont
  {Singh}}\ and\ \bibinfo {author} {\bibfnamefont {L.}~\bibnamefont
  {Nordstr{\"{o}}m}},\ }\href@noop {} {\emph {\bibinfo {title} {Planewaves,
  Pseudopotentials and the LAPW Method, 2nd ed.}}}\ (\bibinfo  {publisher}
  {Springer},\ \bibinfo {address} {Berlin},\ \bibinfo {year}
  {2006})\BibitemShut {NoStop}%
\bibitem [{\citenamefont {Bl\"{o}chl}(1994)}]{BlochlPRB94b}%
  \BibitemOpen
  \bibfield  {author} {\bibinfo {author} {\bibfnamefont {P.~E.}\ \bibnamefont
  {Bl\"{o}chl}},\ }\href@noop {} {\bibfield  {journal} {\bibinfo  {journal}
  {Phys. Rev. B}\ }\textbf {\bibinfo {volume} {50}},\ \bibinfo {pages} {17953}
  (\bibinfo {year} {1994})}\BibitemShut {NoStop}%
\bibitem [{\citenamefont {Sj{\"{o}}stedt}\ \emph {et~al.}(2000)\citenamefont
  {Sj{\"{o}}stedt}, \citenamefont {Nordstr{\"{o}}m},\ and\ \citenamefont
  {Singh}}]{SjostedtSSC00}%
  \BibitemOpen
  \bibfield  {author} {\bibinfo {author} {\bibfnamefont {E.}~\bibnamefont
  {Sj{\"{o}}stedt}}, \bibinfo {author} {\bibfnamefont {L.}~\bibnamefont
  {Nordstr{\"{o}}m}}, \ and\ \bibinfo {author} {\bibfnamefont {D.~J.}\
  \bibnamefont {Singh}},\ }\href@noop {} {\bibfield  {journal} {\bibinfo
  {journal} {Solid State Commun.}\ }\textbf {\bibinfo {volume} {114}},\
  \bibinfo {pages} {15} (\bibinfo {year} {2000})}\BibitemShut {NoStop}%
\bibitem [{\citenamefont {Michalicek}\ \emph {et~al.}(2013)\citenamefont
  {Michalicek}, \citenamefont {Betzinger}, \citenamefont {Friedrich},\ and\
  \citenamefont {Bl\"{u}gel}}]{MichalicekCPC13}%
  \BibitemOpen
  \bibfield  {author} {\bibinfo {author} {\bibfnamefont {G.}~\bibnamefont
  {Michalicek}}, \bibinfo {author} {\bibfnamefont {M.}~\bibnamefont
  {Betzinger}}, \bibinfo {author} {\bibfnamefont {C.}~\bibnamefont
  {Friedrich}}, \ and\ \bibinfo {author} {\bibfnamefont {S.}~\bibnamefont
  {Bl\"{u}gel}},\ }\href@noop {} {\bibfield  {journal} {\bibinfo  {journal}
  {Comput. Phys. Commun.}\ }\textbf {\bibinfo {volume} {184}},\ \bibinfo
  {pages} {2670} (\bibinfo {year} {2013})}\BibitemShut {NoStop}%
\bibitem [{\citenamefont {Wu}\ and\ \citenamefont {Gygi}(2012)}]{WuJCP12}%
  \BibitemOpen
  \bibfield  {author} {\bibinfo {author} {\bibfnamefont {J.}~\bibnamefont
  {Wu}}\ and\ \bibinfo {author} {\bibfnamefont {F.}~\bibnamefont {Gygi}},\
  }\href@noop {} {\bibfield  {journal} {\bibinfo  {journal} {J. Chem. Phys.}\
  }\textbf {\bibinfo {volume} {136}},\ \bibinfo {pages} {224107} (\bibinfo
  {year} {2012})}\BibitemShut {NoStop}%
\bibitem [{\citenamefont {Corsetti}\ \emph {et~al.}(2013)\citenamefont
  {Corsetti}, \citenamefont {Artacho}, \citenamefont {Soler}, \citenamefont
  {Alexandre},\ and\ \citenamefont {Fern\'{a}ndez-Serra}}]{CorsettiJCP13}%
  \BibitemOpen
  \bibfield  {author} {\bibinfo {author} {\bibfnamefont {F.}~\bibnamefont
  {Corsetti}}, \bibinfo {author} {\bibfnamefont {E.}~\bibnamefont {Artacho}},
  \bibinfo {author} {\bibfnamefont {J.~M.}\ \bibnamefont {Soler}}, \bibinfo
  {author} {\bibfnamefont {S.~S.}\ \bibnamefont {Alexandre}}, \ and\ \bibinfo
  {author} {\bibfnamefont {M.-V.}\ \bibnamefont {Fern\'{a}ndez-Serra}},\
  }\href@noop {} {\bibfield  {journal} {\bibinfo  {journal} {J. Chem. Phys.}\
  }\textbf {\bibinfo {volume} {139}},\ \bibinfo {pages} {194502} (\bibinfo
  {year} {2013})}\BibitemShut {NoStop}%
\bibitem [{\citenamefont {Obata}\ \emph {et~al.}(2013)\citenamefont {Obata},
  \citenamefont {Nakamura}, \citenamefont {Hamada},\ and\ \citenamefont
  {Oda}}]{ObataJPSJ13}%
  \BibitemOpen
  \bibfield  {author} {\bibinfo {author} {\bibfnamefont {M.}~\bibnamefont
  {Obata}}, \bibinfo {author} {\bibfnamefont {M.}~\bibnamefont {Nakamura}},
  \bibinfo {author} {\bibfnamefont {I.}~\bibnamefont {Hamada}}, \ and\ \bibinfo
  {author} {\bibfnamefont {T.}~\bibnamefont {Oda}},\ }\href@noop {} {\bibfield
  {journal} {\bibinfo  {journal} {J. Phys. Soc. Jpn.}\ }\textbf {\bibinfo
  {volume} {82}},\ \bibinfo {pages} {093701} (\bibinfo {year}
  {2013})}\BibitemShut {NoStop}%
\bibitem [{\citenamefont {Thonhauser}\ \emph {et~al.}(2015)\citenamefont
  {Thonhauser}, \citenamefont {Zuluaga}, \citenamefont {Arter}, \citenamefont
  {Berland}, \citenamefont {Schr\"oder},\ and\ \citenamefont
  {Hyldgaard}}]{ThonhauserPRL15}%
  \BibitemOpen
  \bibfield  {author} {\bibinfo {author} {\bibfnamefont {T.}~\bibnamefont
  {Thonhauser}}, \bibinfo {author} {\bibfnamefont {S.}~\bibnamefont {Zuluaga}},
  \bibinfo {author} {\bibfnamefont {C.~A.}\ \bibnamefont {Arter}}, \bibinfo
  {author} {\bibfnamefont {K.}~\bibnamefont {Berland}}, \bibinfo {author}
  {\bibfnamefont {E.}~\bibnamefont {Schr\"oder}}, \ and\ \bibinfo {author}
  {\bibfnamefont {P.}~\bibnamefont {Hyldgaard}},\ }\href@noop {} {\bibfield
  {journal} {\bibinfo  {journal} {Phys. Rev. Lett.}\ }\textbf {\bibinfo
  {volume} {115}},\ \bibinfo {pages} {136402} (\bibinfo {year}
  {2015})}\BibitemShut {NoStop}%
\bibitem [{\citenamefont {Schimka}\ \emph {et~al.}(2013)\citenamefont
  {Schimka}, \citenamefont {Gaudoin}, \citenamefont {Klime{\v{s}}},
  \citenamefont {Marsman},\ and\ \citenamefont {Kresse}}]{SchimkaPRB13}%
  \BibitemOpen
  \bibfield  {author} {\bibinfo {author} {\bibfnamefont {L.}~\bibnamefont
  {Schimka}}, \bibinfo {author} {\bibfnamefont {R.}~\bibnamefont {Gaudoin}},
  \bibinfo {author} {\bibfnamefont {J.}~\bibnamefont {Klime{\v{s}}}}, \bibinfo
  {author} {\bibfnamefont {M.}~\bibnamefont {Marsman}}, \ and\ \bibinfo
  {author} {\bibfnamefont {G.}~\bibnamefont {Kresse}},\ }\href@noop {}
  {\bibfield  {journal} {\bibinfo  {journal} {Phys. Rev. B}\ }\textbf {\bibinfo
  {volume} {87}},\ \bibinfo {pages} {214102} (\bibinfo {year}
  {2013})}\BibitemShut {NoStop}%
\bibitem [{\citenamefont {Park}\ \emph {et~al.}(2015)\citenamefont {Park},
  \citenamefont {Yu},\ and\ \citenamefont {Hong}}]{ParkCAP15}%
  \BibitemOpen
  \bibfield  {author} {\bibinfo {author} {\bibfnamefont {J.}~\bibnamefont
  {Park}}, \bibinfo {author} {\bibfnamefont {B.~D.}\ \bibnamefont {Yu}}, \ and\
  \bibinfo {author} {\bibfnamefont {S.}~\bibnamefont {Hong}},\ }\href@noop {}
  {\bibfield  {journal} {\bibinfo  {journal} {Curr. Appl. Phys.}\ }\textbf
  {\bibinfo {volume} {15}},\ \bibinfo {pages} {885} (\bibinfo {year}
  {2015})}\BibitemShut {NoStop}%
\bibitem [{\citenamefont {Blaha}\ \emph {et~al.}(2001)\citenamefont {Blaha},
  \citenamefont {Schwarz}, \citenamefont {Madsen}, \citenamefont {Kvasnicka},\
  and\ \citenamefont {Luitz}}]{WIEN2k}%
  \BibitemOpen
  \bibfield  {author} {\bibinfo {author} {\bibfnamefont {P.}~\bibnamefont
  {Blaha}}, \bibinfo {author} {\bibfnamefont {K.}~\bibnamefont {Schwarz}},
  \bibinfo {author} {\bibfnamefont {G.~K.~H.}\ \bibnamefont {Madsen}}, \bibinfo
  {author} {\bibfnamefont {D.}~\bibnamefont {Kvasnicka}}, \ and\ \bibinfo
  {author} {\bibfnamefont {J.}~\bibnamefont {Luitz}},\ }\href@noop {} {\emph
  {\bibinfo {title} {WIEN2K: An Augmented Plane Wave plus Local Orbitals
  Program for Calculating Crystal Properties}}}\ (\bibinfo  {publisher} {Vienna
  University of Technology},\ \bibinfo {address} {Austria},\ \bibinfo {year}
  {2001})\BibitemShut {NoStop}%
\bibitem [{\citenamefont {Giannozzi}\ \emph {et~al.}(2009)\citenamefont
  {Giannozzi}, \citenamefont {Baroni}, \citenamefont {Bonini}, \citenamefont
  {Calandra}, \citenamefont {Car}, \citenamefont {Cavazzoni}, \citenamefont
  {Ceresoli}, \citenamefont {Chiarotti}, \citenamefont {Cococcioni},
  \citenamefont {Dabo}, \citenamefont {Dal~Corso}, \citenamefont
  {de~Gironcoli}, \citenamefont {Fabris}, \citenamefont {Fratesi},
  \citenamefont {Gebauer}, \citenamefont {Gerstmann}, \citenamefont
  {Gougoussis}, \citenamefont {Kokalj}, \citenamefont {Lazzeri}, \citenamefont
  {Martin-Samos}, \citenamefont {Marzari}, \citenamefont {Mauri}, \citenamefont
  {Mazzarello}, \citenamefont {Paolini}, \citenamefont {Pasquarello},
  \citenamefont {Paulatto}, \citenamefont {Sbraccia}, \citenamefont {Scandolo},
  \citenamefont {Sclauzero}, \citenamefont {Seitsonen}, \citenamefont
  {Smogunov}, \citenamefont {Umari},\ and\ \citenamefont
  {Wentzcovitch}}]{GiannozziJPCM09}%
  \BibitemOpen
  \bibfield  {author} {\bibinfo {author} {\bibfnamefont {P.}~\bibnamefont
  {Giannozzi}}, \bibinfo {author} {\bibfnamefont {S.}~\bibnamefont {Baroni}},
  \bibinfo {author} {\bibfnamefont {N.}~\bibnamefont {Bonini}}, \bibinfo
  {author} {\bibfnamefont {M.}~\bibnamefont {Calandra}}, \bibinfo {author}
  {\bibfnamefont {R.}~\bibnamefont {Car}}, \bibinfo {author} {\bibfnamefont
  {C.}~\bibnamefont {Cavazzoni}}, \bibinfo {author} {\bibfnamefont
  {D.}~\bibnamefont {Ceresoli}}, \bibinfo {author} {\bibfnamefont {G.~L.}\
  \bibnamefont {Chiarotti}}, \bibinfo {author} {\bibfnamefont {M.}~\bibnamefont
  {Cococcioni}}, \bibinfo {author} {\bibfnamefont {I.}~\bibnamefont {Dabo}},
  \bibinfo {author} {\bibfnamefont {A.}~\bibnamefont {Dal~Corso}}, \bibinfo
  {author} {\bibfnamefont {S.}~\bibnamefont {de~Gironcoli}}, \bibinfo {author}
  {\bibfnamefont {S.}~\bibnamefont {Fabris}}, \bibinfo {author} {\bibfnamefont
  {G.}~\bibnamefont {Fratesi}}, \bibinfo {author} {\bibfnamefont
  {R.}~\bibnamefont {Gebauer}}, \bibinfo {author} {\bibfnamefont
  {U.}~\bibnamefont {Gerstmann}}, \bibinfo {author} {\bibfnamefont
  {C.}~\bibnamefont {Gougoussis}}, \bibinfo {author} {\bibfnamefont
  {A.}~\bibnamefont {Kokalj}}, \bibinfo {author} {\bibfnamefont
  {M.}~\bibnamefont {Lazzeri}}, \bibinfo {author} {\bibfnamefont
  {L.}~\bibnamefont {Martin-Samos}}, \bibinfo {author} {\bibfnamefont
  {N.}~\bibnamefont {Marzari}}, \bibinfo {author} {\bibfnamefont
  {F.}~\bibnamefont {Mauri}}, \bibinfo {author} {\bibfnamefont
  {R.}~\bibnamefont {Mazzarello}}, \bibinfo {author} {\bibfnamefont
  {S.}~\bibnamefont {Paolini}}, \bibinfo {author} {\bibfnamefont
  {A.}~\bibnamefont {Pasquarello}}, \bibinfo {author} {\bibfnamefont
  {L.}~\bibnamefont {Paulatto}}, \bibinfo {author} {\bibfnamefont
  {C.}~\bibnamefont {Sbraccia}}, \bibinfo {author} {\bibfnamefont
  {S.}~\bibnamefont {Scandolo}}, \bibinfo {author} {\bibfnamefont
  {G.}~\bibnamefont {Sclauzero}}, \bibinfo {author} {\bibfnamefont {A.~P.}\
  \bibnamefont {Seitsonen}}, \bibinfo {author} {\bibfnamefont {A.}~\bibnamefont
  {Smogunov}}, \bibinfo {author} {\bibfnamefont {P.}~\bibnamefont {Umari}}, \
  and\ \bibinfo {author} {\bibfnamefont {R.~M.}\ \bibnamefont {Wentzcovitch}},\
  }\href@noop {} {\bibfield  {journal} {\bibinfo  {journal} {J. Phys.: Condens.
  Matter}\ }\textbf {\bibinfo {volume} {21}},\ \bibinfo {pages} {395502}
  (\bibinfo {year} {2009})}\BibitemShut {NoStop}%
\bibitem [{FFT()}]{FFTW}%
  \BibitemOpen
  \href@noop {} {}\bibinfo {howpublished} {See http://www.fftw.org}\BibitemShut
  {NoStop}%
\bibitem [{\citenamefont {Tao}\ \emph {et~al.}(2010)\citenamefont {Tao},
  \citenamefont {Perdew},\ and\ \citenamefont {Ruzsinszky}}]{TaoPRB10}%
  \BibitemOpen
  \bibfield  {author} {\bibinfo {author} {\bibfnamefont {J.}~\bibnamefont
  {Tao}}, \bibinfo {author} {\bibfnamefont {J.~P.}\ \bibnamefont {Perdew}}, \
  and\ \bibinfo {author} {\bibfnamefont {A.}~\bibnamefont {Ruzsinszky}},\
  }\href@noop {} {\bibfield  {journal} {\bibinfo  {journal} {Phys. Rev. B}\
  }\textbf {\bibinfo {volume} {81}},\ \bibinfo {pages} {233102} (\bibinfo
  {year} {2010})}\BibitemShut {NoStop}%
\bibitem [{\citenamefont {Sorescu}\ \emph {et~al.}(2014)\citenamefont
  {Sorescu}, \citenamefont {Byrd}, \citenamefont {Rice},\ and\ \citenamefont
  {Jordan}}]{SorescuJCTC14}%
  \BibitemOpen
  \bibfield  {author} {\bibinfo {author} {\bibfnamefont {D.~C.}\ \bibnamefont
  {Sorescu}}, \bibinfo {author} {\bibfnamefont {E.~F.~C.}\ \bibnamefont
  {Byrd}}, \bibinfo {author} {\bibfnamefont {B.~M.}\ \bibnamefont {Rice}}, \
  and\ \bibinfo {author} {\bibfnamefont {K.~D.}\ \bibnamefont {Jordan}},\
  }\href@noop {} {\bibfield  {journal} {\bibinfo  {journal} {J. Chem. Theory
  Comput.}\ }\textbf {\bibinfo {volume} {10}},\ \bibinfo {pages} {4982}
  (\bibinfo {year} {2014})}\BibitemShut {NoStop}%
\bibitem [{\citenamefont {Callsen}\ and\ \citenamefont
  {Hamada}(2015)}]{CallsenPRB15}%
  \BibitemOpen
  \bibfield  {author} {\bibinfo {author} {\bibfnamefont {M.}~\bibnamefont
  {Callsen}}\ and\ \bibinfo {author} {\bibfnamefont {I.}~\bibnamefont
  {Hamada}},\ }\href@noop {} {\bibfield  {journal} {\bibinfo  {journal} {Phys.
  Rev. B}\ }\textbf {\bibinfo {volume} {91}},\ \bibinfo {pages} {195103}
  (\bibinfo {year} {2015})},\ \bibinfo {note} {\textbf{95}, 039905(E)
  (2017)}\BibitemShut {NoStop}%
\bibitem [{\citenamefont {Graziano}\ \emph {et~al.}(2012)\citenamefont
  {Graziano}, \citenamefont {Klime{\v{s}}}, \citenamefont {Fernandez-Alonso},\
  and\ \citenamefont {Michaelides}}]{GrazianoJPCM12}%
  \BibitemOpen
  \bibfield  {author} {\bibinfo {author} {\bibfnamefont {G.}~\bibnamefont
  {Graziano}}, \bibinfo {author} {\bibfnamefont {J.}~\bibnamefont
  {Klime{\v{s}}}}, \bibinfo {author} {\bibfnamefont {F.}~\bibnamefont
  {Fernandez-Alonso}}, \ and\ \bibinfo {author} {\bibfnamefont
  {A.}~\bibnamefont {Michaelides}},\ }\href@noop {} {\bibfield  {journal}
  {\bibinfo  {journal} {J. Phys.: Condens. Matter}\ }\textbf {\bibinfo {volume}
  {24}},\ \bibinfo {pages} {424216} (\bibinfo {year} {2012})}\BibitemShut
  {NoStop}%
\bibitem [{\citenamefont {Bj\"orkman}\ \emph {et~al.}(2012)\citenamefont
  {Bj\"orkman}, \citenamefont {Gulans}, \citenamefont {Krasheninnikov},\ and\
  \citenamefont {Nieminen}}]{BjorkmanPRL12}%
  \BibitemOpen
  \bibfield  {author} {\bibinfo {author} {\bibfnamefont {T.}~\bibnamefont
  {Bj\"orkman}}, \bibinfo {author} {\bibfnamefont {A.}~\bibnamefont {Gulans}},
  \bibinfo {author} {\bibfnamefont {A.~V.}\ \bibnamefont {Krasheninnikov}}, \
  and\ \bibinfo {author} {\bibfnamefont {R.~M.}\ \bibnamefont {Nieminen}},\
  }\href@noop {} {\bibfield  {journal} {\bibinfo  {journal} {Phys. Rev. Lett.}\
  }\textbf {\bibinfo {volume} {108}},\ \bibinfo {pages} {235502} (\bibinfo
  {year} {2012})}\BibitemShut {NoStop}%
\bibitem [{\citenamefont {Bj\"orkman}(2014)}]{BjorkmanJCP14}%
  \BibitemOpen
  \bibfield  {author} {\bibinfo {author} {\bibfnamefont {T.}~\bibnamefont
  {Bj\"orkman}},\ }\href@noop {} {\bibfield  {journal} {\bibinfo  {journal} {J.
  Chem. Phys.}\ }\textbf {\bibinfo {volume} {141}},\ \bibinfo {pages} {074708}
  (\bibinfo {year} {2014})}\BibitemShut {NoStop}%
\bibitem [{\citenamefont {Kresse}\ and\ \citenamefont
  {Furthm\"uller}(1996)}]{KressePRB96}%
  \BibitemOpen
  \bibfield  {author} {\bibinfo {author} {\bibfnamefont {G.}~\bibnamefont
  {Kresse}}\ and\ \bibinfo {author} {\bibfnamefont {J.}~\bibnamefont
  {Furthm\"uller}},\ }\href@noop {} {\bibfield  {journal} {\bibinfo  {journal}
  {Phys. Rev. B}\ }\textbf {\bibinfo {volume} {54}},\ \bibinfo {pages} {11169}
  (\bibinfo {year} {1996})}\BibitemShut {NoStop}%
\bibitem [{\citenamefont {VandeVondele}\ \emph {et~al.}(2005)\citenamefont
  {VandeVondele}, \citenamefont {Krack}, \citenamefont {Mohamed}, \citenamefont
  {Parrinello}, \citenamefont {Chassaing},\ and\ \citenamefont
  {Hutter}}]{VandeVondeleCPC05}%
  \BibitemOpen
  \bibfield  {author} {\bibinfo {author} {\bibfnamefont {J.}~\bibnamefont
  {VandeVondele}}, \bibinfo {author} {\bibfnamefont {M.}~\bibnamefont {Krack}},
  \bibinfo {author} {\bibfnamefont {F.}~\bibnamefont {Mohamed}}, \bibinfo
  {author} {\bibfnamefont {M.}~\bibnamefont {Parrinello}}, \bibinfo {author}
  {\bibfnamefont {T.}~\bibnamefont {Chassaing}}, \ and\ \bibinfo {author}
  {\bibfnamefont {J.}~\bibnamefont {Hutter}},\ }\href@noop {} {\bibfield
  {journal} {\bibinfo  {journal} {Comput. Phys. Commun.}\ }\textbf {\bibinfo
  {volume} {167}},\ \bibinfo {pages} {103} (\bibinfo {year}
  {2005})}\BibitemShut {NoStop}%
\bibitem [{\citenamefont {Perdew}\ \emph {et~al.}(1996)\citenamefont {Perdew},
  \citenamefont {Burke},\ and\ \citenamefont {Ernzerhof}}]{PerdewPRL96}%
  \BibitemOpen
  \bibfield  {author} {\bibinfo {author} {\bibfnamefont {J.~P.}\ \bibnamefont
  {Perdew}}, \bibinfo {author} {\bibfnamefont {K.}~\bibnamefont {Burke}}, \
  and\ \bibinfo {author} {\bibfnamefont {M.}~\bibnamefont {Ernzerhof}},\
  }\href@noop {} {\bibfield  {journal} {\bibinfo  {journal} {Phys. Rev. Lett.}\
  }\textbf {\bibinfo {volume} {77}},\ \bibinfo {pages} {3865} (\bibinfo {year}
  {1996})};\ \bibinfo {note} {\textbf{78}, 1396(E) (1997)}\BibitemShut
  {NoStop}%
\bibitem [{\citenamefont {Bj\"orkman}(2012)}]{BjorkmanPRB12}%
  \BibitemOpen
  \bibfield  {author} {\bibinfo {author} {\bibfnamefont {T.}~\bibnamefont
  {Bj\"orkman}},\ }\href@noop {} {\bibfield  {journal} {\bibinfo  {journal}
  {Phys. Rev. B}\ }\textbf {\bibinfo {volume} {86}},\ \bibinfo {pages} {165109}
  (\bibinfo {year} {2012})}\BibitemShut {NoStop}%
\bibitem [{\citenamefont {Weinert}(1981)}]{WeinertJMP81}%
  \BibitemOpen
  \bibfield  {author} {\bibinfo {author} {\bibfnamefont {M.}~\bibnamefont
  {Weinert}},\ }\href@noop {} {\bibfield  {journal} {\bibinfo  {journal} {J.
  Math. Phys.}\ }\textbf {\bibinfo {volume} {22}},\ \bibinfo {pages} {2433}
  (\bibinfo {year} {1981})}\BibitemShut {NoStop}%
\bibitem [{\citenamefont {Kara}\ and\ \citenamefont
  {Kurki-Suonio}(1981)}]{KaraAC81}%
  \BibitemOpen
  \bibfield  {author} {\bibinfo {author} {\bibfnamefont {M.}~\bibnamefont
  {Kara}}\ and\ \bibinfo {author} {\bibfnamefont {K.}~\bibnamefont
  {Kurki-Suonio}},\ }\href@noop {} {\bibfield  {journal} {\bibinfo  {journal}
  {Acta Cryst.}\ }\textbf {\bibinfo {volume} {A37}},\ \bibinfo {pages} {201}
  (\bibinfo {year} {1981})}\BibitemShut {NoStop}%
\bibitem [{\citenamefont {Arfken}\ and\ \citenamefont {Weber}(2005)}]{Arfken}%
  \BibitemOpen
  \bibfield  {author} {\bibinfo {author} {\bibfnamefont {G.~B.}\ \bibnamefont
  {Arfken}}\ and\ \bibinfo {author} {\bibfnamefont {H.~J.}\ \bibnamefont
  {Weber}},\ }\href@noop {} {\emph {\bibinfo {title} {Mathematical Methods for
  Physicists, 6th ed.}}}\ (\bibinfo  {publisher} {Elsevier Academic Press},\
  \bibinfo {address} {San Diego, CA},\ \bibinfo {year} {2005})\BibitemShut
  {NoStop}%
\end{thebibliography}%

\end{document}